%% file: main.tex
\documentclass[prx,aps,reprint,showpacs,amsfonts,amsmath,floatfix,superscriptaddress,footinbib]{revtex4-2}

\usepackage[svgnames]{xcolor}

\usepackage[colorlinks=true,        
            allcolors = black,  
            citecolor=DarkBlue, 
            linkcolor=DarkBlue,
            urlcolor=DarkBlue
        ]{hyperref} 

\usepackage{amsmath, amsfonts, amssymb, amsthm, bbm}
\usepackage{dsfont}
\usepackage{bm} 
\usepackage{mathtools}
\usepackage{physics}
\usepackage{graphicx}
\usepackage{dcolumn}

\usepackage{tikz}
\usetikzlibrary{quantikz}
\usepackage{graphicx}
\usepackage{dcolumn}

\usepackage{rotating}

\DeclarePairedDelimiter\ceil{\lceil}{\rceil} 
\DeclarePairedDelimiter{\floor}{\lfloor}{\rfloor} 

\usepackage{array}   
\newcolumntype{L}{>{$}l<{$}} 

\newcommand{\pauli}[2]{\sigma_{#1}^{#2}} 
\newcommand{\cardvec}{\mathbf{C}}
\newcommand{\cardelement}{c}
\newcommand{\numqubits}{M}
\newcommand{\numterms}{K}
\newcommand{\numaminoacids}{N}

\newcommand{\bubin}{\text{BUBinary}}
\newcommand{\oh}{\text{Unary}}
\newcommand{\binary}{\text{Binary}}

\renewcommand{\selectlanguage}[1]{} 

\usepackage{comment} 

\begin{document}

\preprint{APS/123-QED}

\title{Resource analysis of quantum algorithms for coarse-grained protein folding models}

\author{Hanna Linn}
\email{hannlinn@chalmers.se}
\author{Isak Brundin}%
\author{Laura García-Álvarez}
\author{G\"{o}ran Johansson}
\affiliation{%
Department of Microtechnology and Nanoscience (MC2), Chalmers University of Technology, SE-412 96 G\"{o}teborg, Sweden
}%

\begin{abstract}
Protein folding processes are a vital aspect of molecular biology that is hard to simulate with conventional computers. Quantum algorithms have been proven superior for certain problems and may help tackle this complex life science challenge.
We analyze the resource requirements for simulating protein folding on a quantum computer, assessing this problem's feasibility in the current and near-future technological landscape. We calculate the minimum number of qubits, interactions, and two-qubit gates necessary to build a heuristic quantum algorithm with the specific information of a folding problem.
Particularly, we focus on the resources needed to build quantum operations based on the Hamiltonian linked to the protein folding models for a given amino acid count. Such operations are a fundamental component of these quantum algorithms, guiding the evolution of the quantum state for efficient computations.
Specifically, we study course-grained folding models on the lattice and the fixed backbone side-chain conformation model and assess their compatibility with the constraints of existing quantum hardware given different bit-encodings.
We conclude that the number of qubits required falls within current technological capabilities. However, the limiting factor is the high number of interactions in the Hamiltonian, resulting in a quantum gate count unavailable today.
\end{abstract}

\maketitle


\section{Introduction}
\label{sec:iso} 
Shor's and Grover's algorithms demonstrate exponential and polynomial speedups compared to classical methods for the practical problems of prime factorization~\cite{shor_polynomial-time_1997} and unsorted searching~\cite{grover_fast_1996}. This breakthrough ignited the pursuit of further quantum algorithms and use cases, aiming to achieve quantum speedup in computations for real-world applications. However, reaching such an advantage typically requires fully error-corrected devices yet to be available, with around $10^5$ qubits~\cite{jones_layered_2012}.
Until such platforms become a reality, the community explores algorithms suitable for the current Noisy Intermediate-Scale Quantum (NISQ) computers~\cite{preskill_quantum_2018}, with fifty to a few hundred noisy qubits. The current efforts to realize a quantum computer involve various approaches, such as superconducting qubits~\cite{kjaergaard_superconducting_2020}, trapped ion qubits~\cite{bruzewicz_trapped-ion_2019}, and photonic qubits~\cite{zhong_quantum_2020}, each with its technical challenges and trade-offs regarding qubit count, connectivity, and coherence time. The latter limits the gate fidelities so that the computation will be dominated by noise, usually well before performing even a thousand gates on each qubit.

In the NISQ era, hybrid quantum-classical algorithms have attracted considerable attention for their ability to harness existing hardware capabilities and potentially provide a quantum advantage for specific computational problems~\cite{cerezo_variational_2021}, with one notable example being the Quantum Approximate Optimization Algorithm (QAOA)~\cite{farhi_quantum_2014}.
A customized variational ansatz containing solutions to a problem depends on a finite number of classically optimizable parameters related to a heuristic quantum algorithm. This ansatz construction commonly involves using problem-specific information, such as the energy or cost function of a given optimization problem.
Hybrid algorithms are highly adaptable and thereby find applications in various domains, including chemistry~\cite{kandala_hardware-efficient_2017}, machine learning~\cite{caunhye_optimization_2012}, and optimization~\cite{farhi_quantum_2014}, as well as applications in the field of protein folding~\cite{fingerhuth_quantum_2018, robert_resource-efficient_2021}.

Studying protein folding is essential for understanding how proteins gain their functional three-dimensional structures, comprehending the protein's biological functions, and designing effective therapeutic interventions.
The folding from an amino acid sequence to a stable structure is deeply intricate~\cite{levinthal_are_1968,zwanzig_levinthals_1992}, and simulating the dynamics represents a complex optimization problem where one wants to find the conformation with the lowest score according to an energy function~\cite{piana_assessing_2014, dill_protein_2008}.
The field of in silico protein structure prediction has seen massive success recently with classical heuristics~\cite{jumper_highly_2021, kuhlman_advances_2019, dorn_three-dimensional_2014}. Before the availability of computational resources enabling these fully atomistic predictions, course-graining models served as an intermediary step and continue to be essential in multiscale modeling today.
Course-grained models decrease the sampling space and lower the complexity of the problem by assuming various levels of reduced polypeptide chain representation~\cite{kmiecik_coarse-grained_2016}. Similarly, due to the limited quantum resources currently available, quantum algorithms must consider these simplifications. We focus on the course-grained models previously explored in the field of quantum algorithms to give an overview of the field: the lattice model~\cite{levitt_computer_1975,hinds_lattice_1992} with hydrophobic-polar (HP) energies~\cite{lau_lattice_1989} or Miyazawa-Jernigan (MJ) energies~\cite{miyazawa_residue_1996}, and an off-lattice side-chain conformation based model with a fixed backbone~\cite{jacobson_force_2002, bower_prediction_1997}.

The first proof-of-concept proposals for finding optimal protein folds with quantum devices consider only small peptides due to the quantum hardware limitations. 
Early formulations for quantum annealers include the HP-lattice model on a square grid, with the amino acid coordinates encoded with binary strings~\cite{perdomo_construction_2008}. The high resource demand of this initial approach motivated the search for new problem formulations, leading to a divide-and-conquer method and a turn-based encoding of the protein conformation. Consequently, a quantum annealer successfully found the optimal fold of a chain with six amino acids for a model with MJ energies~\cite{perdomo-ortiz_finding_2012}, and this formulation was adapted to QAOA for an ion-trap experiment~\cite{saito_lattice_2023}. Subsequent works lowered the quantum operations needed~\cite{babbush_construction_2014}, which prompted further experiments on a quantum annealer tackling the folding of a ten-amino acid chain on a planar lattice and an eight-amino acid chain on a cubic lattice~\cite{babej_coarse-grained_2018}.
The quantum algorithm improvements for constrained optimization problems resulted in the use of the Quantum Alternating Operator Ansatz~\cite{hadfield_quantum_2019} to fold a four amino acid long chain on an ion-trap quantum computer and further explorations to incorporate efficiently the problem constraints in the algorithm formulation~\cite{fingerhuth_quantum_2018}.
Such strategic engineering of the problem formulation can also improve the success of the folding problem on quantum annealers~\cite{outeiral_investigating_2021}.
Moreover, the field benefited from new variational quantum algorithms such as the Conditional Value-at-Risk (CVaR)-Variational Quantum Eigensolver~\cite{barkoutsos_improving_2020}. This algorithm, combined with a new resource-efficient model with MJ energies and a tetrahedral lattice, was implemented experimentally on a superconducting circuit quantum computer for a seven-amino acid long chain~\cite{robert_resource-efficient_2021}.
Despite the fact that numerical simulations of QAOA for this tetrahedral lattice model show lower performance, even for a sequence of four amino acids~\cite{boulebnane_peptide_2022}, when combining this model with other algorithms such as the Digitized-Counterdiabatic Quantum Algorithm, it succeeded in folding a nine-amino acid chain on two superconducting circuit gate-based quantum devices and an ion-trap platform~\cite{chandarana_digitized_2023}.
Likewise, parallel efforts led to the realization of an HP-lattice model for a fourteen-amino acid long chain on a quantum annealer. The model reduced the qubits needed for a coordinate-based conformation encoding by including lattice symmetries~\cite{irback_folding_2022}.
In these early stages, the simplicity of the HP-lattice model compared to higher-resolution descriptions positions it among the central models for different quantum algorithms, including Grover-based protocols~\cite{wong_fast_2022}.
Recently, the field has also expanded into off-lattice models. These works range from formulations of peptide packing for a quantum annealer using side-chain conformation-based models~\cite{mulligan_designing_2020,maguire_xenet_2021} to methods using deep learning for initial state generation followed by a quantum Metropolis-Hastings algorithm to decide parameterized torsion angles of a tetrapeptide~\cite{casares_qfold_2022}.

In every previous formulation, the authors have made different choices regarding the protein folding model and its encoding into the quantum computer. The models' resolution and translation to quantum variables impact the algorithm's time- and space-complexity. This translation includes representing a specific folding or conformation with qubits and expressing the model's energy function using interactions between them.
In essence, the resulting problem formulation as a quantum spin model includes the interactions between the course-grained beads of the protein and the folding constraints in the so-called cost Hamiltonian.
We can customize this Hamiltonian to focus on specific aspects of the problem or exploit particular symmetries or properties.
The size and complexity of the cost Hamiltonian will affect the corresponding required quantum circuit. That is, it impacts the run time or circuit depth, the number of qubits, and how they interact, which will, in turn, influence the choice of hardware.
Optimal decisions in the encoding step can reduce the need for quantum hardware resources, while suboptimal ones may strain connectivities and operations allowed in the devices. Due to the diverse capabilities inherent in quantum platforms~\cite{blinov_comparison_2021}, resource trade-offs arise contingent upon the chosen encoding strategy.

Improving quantum algorithms requires analyzing how the problem encoding can affect the necessary resources and the demands on current quantum technology. Accordingly, the community has studied encoding resource trade-offs for quantum and discrete optimization problems~\cite{sawaya_resource-efficient_2020,sawaya_encoding_2023}. A more generalized approach includes an extensive library of hardware-independent formulations of these problems for quantum computing given different encodings~\cite{dominguez_encoding-independent_2023}. 
In contrast, another analysis focused on the feasibility of several hybrid quantum-classical algorithms on current quantum computers for Maximum Independent Set---an optimization problem over binary variables. There, as the encodings do not play a role, they detailed the quantum gates and classical resources needed, showcasing trade-offs between gate decomposition methods for different quantum hardware~\cite{tomesh_quantum-classical_2022}.

In this paper, we explore several encodings of discrete optimization formulations of protein folding into quantum variables and analyze the resource requirements given the characteristics of different quantum hardware.
Our objective is to assess the suitability of gate-based quantum computers, specifically NISQ devices, to potentially fold proteins of realistic size.
Human proteins have a median length of 375 amino acids~\cite{brocchieri_protein_2005}, whereas clinical target proteins tend to be about 414 amino acids long on average~\cite{wang_therapeutic_2020}. So, what kind of quantum computer would we need to fold a hundred amino acids? We examine lattice protein folding and side-chain packing models and detail the required quantum resources---the number of qubits,  interactions, and the resulting two-qubit gates---linked to the \oh{}, \binary{}, and \bubin{} encodings.
In particular, we only estimate the resources needed to include the building block operation linked to the cost Hamiltonian in the quantum algorithm. That is, we provide a lower bound of the resources required to address these protein folding models with quantum heuristics for current technology. It remains an open question whether these quantum algorithms can succeed or for what algorithm depth they perform well.

The paper is structured as follows. In Sec.~\ref{sec:Coarse-grained protein models}, we describe coarse-grained protein models and, in particular, the discrete variable formulation of HP-lattice models and side-chain conformation-based models. In Sec.~\ref{sec:DVtoqubits}, we present the \oh{}, \binary{}, and \bubin{} encodings for translating discrete variables into qubits and their application to the previous models. Further, in Sec.~\ref{sec:resource_trade}, we analyze the trade-offs associated with qubit growth, gates, and the size of the unfeasible solution set in the encoding Hamiltonians. Finally, Sec.~\ref{sec:conclusion} concludes the discussion by summarizing the key findings, highlighting their implications, and suggesting future research directions.

\section{Coarse-grained protein models}
\label{sec:Coarse-grained protein models}   
Studying proteins is challenging due to the size of the systems and their interactions. We focus on coarse-grained protein models that lower the degrees of freedom in the polypeptide representation~\cite{kmiecik_coarse-grained_2016}. Different simplifications lead to computationally advantageous low-resolution models that capture different system properties and behaviors, enabling the study of protein folding mechanisms or protein structure prediction, among others.
These low-resolution models characterize larger systems over longer timescales and may be combined with atomistic simulations in multiscale modeling schemes.

We consider coarse-grained models designed to understand the protein folding process and suitable for quantum computing architectures. The models include two principal reductions to ensure their feasibility with the state-of-the-art and near-term quantum technology. Firstly, chains of beads with different properties represent groups of atoms in amino acids. Secondly, we use discrete representations of the structures' geometry---the chain of beads is placed on lattice grids or discretized spatial orientations---such that the finding of the optimal fold can be formulated as a discrete optimization problem.

\begin{figure}[htbp]
    \centering
    \includegraphics[width=0.90\linewidth]{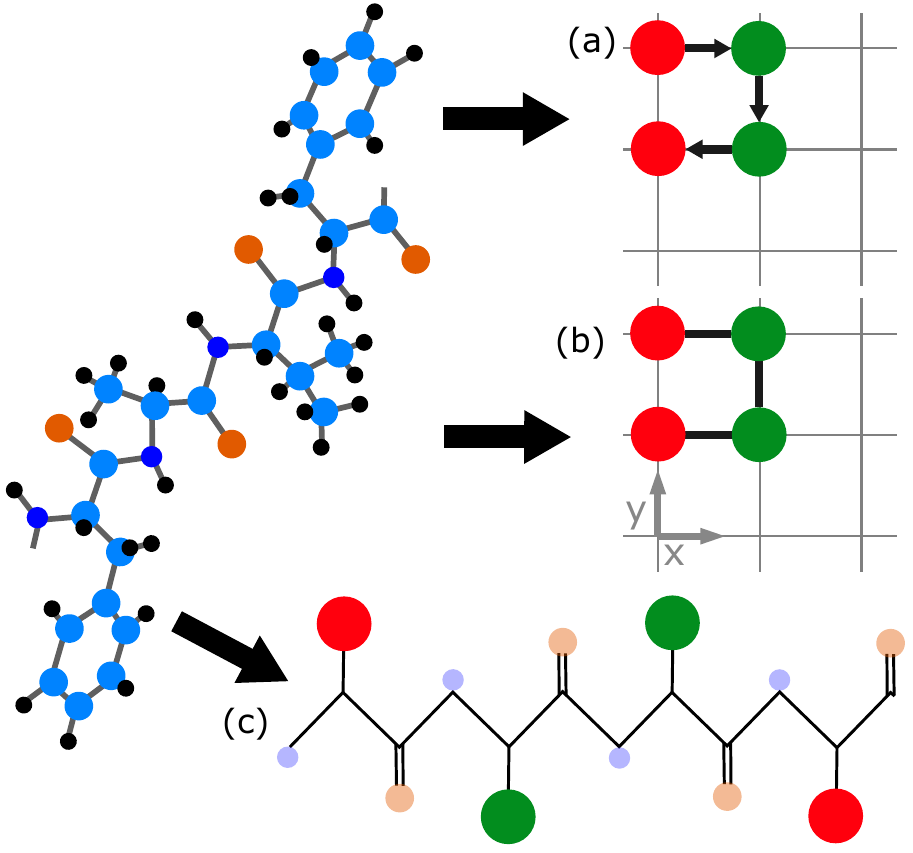}
    \caption{Fully atomistic depiction of a four amino acid chain and the corresponding course-grain model representations for the (a) turn-based lattice model, (b) coordinate-based lattice model, and (c) conformational-based model.}
    \label{fig:coarse-grain}
\end{figure}

The chain representation of the protein varies according to the level of resolution.
Despite the huge simplifications of these lattice models, finding their optimal protein conformations is an NP-hard problem that challenges the capabilities of conventional computers~\cite{berger_protein_1998, hart_robust_1997, fraenkel_complexity_1993, crescenzi_complexity_1998}.
Moreover, other methods replace only partially the amino acid with a bead, such as Rosetta's centroid representation, where the backbone remains atomistic, and the beads replace the side chains.
We consider these approximations and focus on coarse-grained models addressed previously by the quantum computing community, which can be formulated as discrete optimization problems. Namely, lattice models~\cite{perdomo_construction_2008, perdomo-ortiz_finding_2012, babbush_construction_2014, fingerhuth_quantum_2018, babej_coarse-grained_2018, outeiral_investigating_2021, robert_resource-efficient_2021, irback_folding_2022, saito_lattice_2023} and the side-chain conformation-based model~\cite{mulligan_designing_2020,maguire_xenet_2021}. For the sake of simplicity, we consider two-dimensional square and three-dimensional cubic lattices.

To compare the resource requirements---the number of qubits, interactions, and gates---for the quantum hardware implementation of the different models, we use a general formulation that accounts for the number of possible conformations of each model. 
Given a sequence of $N$ amino acids $(a_{1}, \dots, a_{N})$ for which each amino acid $a_i$ can take a conformation from the finite set of integers, $R_i=\{1,2, \dots, \cardelement_i\}$, we define a vector with the cardinalities of the sets of conformations $\cardelement_i=n(R_i)$ as
\begin{equation}
    \cardvec = (\cardelement_1, \cardelement_2, \dots, \cardelement_{N}).
    \label{eq:card_vector}
\end{equation}
Given a choice $r_i \in R_i$ for each amino acid, we create a bitstring $x= r_1r_2\dots r_N$ encoding a protein conformation by concatenating the binary representation of each integer $r_i$.
Various methods exist to convert the integer information into binary strings, presented in Sec.~\ref{sec:DVtoqubits}.
To perform the discrete optimization, we need an energy function that assesses the energy of the bitstring as we want to find the bitstring encoding the protein conformation with the lowest energy. This function combines negative potentials from interacting amino acids with positive penalty terms, maintaining physical constraints.

We use this framework to describe the lattice and side-chain conformation-based models in Secs.~\ref{subsec:HP_models} and \ref{subsec:Conf_model}, respectively. See Fig.~\ref{fig:coarse-grain} for a schematic representation of the coarse-grained simplifications.
The vector $\cardvec$ varies across models. In lattice models, it is insufficient to describe allowed bitstrings entirely. In coordinate-based models, extra frameworks are needed for physical conformation, while turn-based models require additional qubits to assess bead interactions.

\subsection{HP-lattice models}
\label{subsec:HP_models} 

In the HP-lattice model, a protein is represented by a chain $(a_{1}, \dots, a_{\numaminoacids})$ of hydrophobic and polar beads, $a_i \in \{\text{H,P}\}$, placed on a lattice.
This model consists of an energy function whose ground state corresponds to the optimal fold that minimizes the exposure of hydrophobic residues to the solvent while penalizing non-physical configurations.
Firstly, one considers a pairwise potential that depends on the distance between beads and accounts for the number of nearest-neighbor hydrophobic beads in the conformation. Secondly, we impose constraints on the possible folds such that the chain is not overlapping---with two amino acids occupying the same lattice site---or broken.
With lattice grids, the distance in the energy function can be formulated in terms of integer variables, leading to a classical discrete optimization problem.

We study the two most common ways of encoding the sequence placement on the lattice and, therefore, the distance between beads. On the one hand, seminal works in quantum computation use coordinate-based encodings, for which one needed at least $D\numaminoacids \log_2 \numaminoacids$ qubits, with $D$ the lattice dimension and $\numaminoacids$ the number of amino acids~\cite{perdomo_construction_2008}.
On the other hand, further developments introduced a turn-based encoding, in which the eventual translation to quantum hardware reduced the qubit requirements at the cost of more operations~\cite{perdomo-ortiz_finding_2012}.

\paragraph*{Coordinate-based HP-lattice model.}
In this model, the protein conformation is encoded with the lattice coordinates of each bead. That is, the integer components of the vector in Eq.~(\ref{eq:card_vector}) is the number of possible positions of each chain component $a_i$ on the lattice grid. This work focuses on the model presented in Ref.~\cite{irback_folding_2022}.
We consider the additional checkerboard symmetry that divides the number of available positions for each bead in half~\cite{irback_folding_2022}. Essentially, we label the grid locations and chain beads as odd or even, such that each bead can only be placed in a location with the same parity.
In the two-dimensional case, we repeat a square unit cell in two directions to generate a lattice of side lengths $L_1$ and $L_2$. Therefore, the number of possible available conformations for each bead is given by the cardinality vector
\begin{equation}
    \cardvec_{\text{HP-square}}^{\text{coord}} = \left(\ceil*{\frac{L_1L_2}{2}}, \floor*{\frac{L_1L_2}{2}}, \ceil*{\frac{L_1L_2}{2}}, \dots\right).
    \label{eq:encvec_coord_2D}
\end{equation}

We also consider a three-dimensional lattice with side lengths of $L_1$, $L_2$, and $L_3$ created from a cubic unit cell. In this case, the number of possible locations for the beads can take values up to  
\begin{equation}
    \cardvec_{\text{HP-cubic}}^{\text{coord}} = \left(\ceil*{\frac{L_1L_2L_3}{2}}, \floor*{\frac{L_1L_2L_3}{2}}, \ceil*{\frac{L_1L_2L_3}{2}}, \dots\right).
    \label{eq:encvec_coord_3D}
\end{equation}

The energy function also utilizes checkerboard symmetry, evaluating the pairwise potential for adjacent even and odd sites.
Penalty terms are used to enforce three constraints that help to eliminate non-physical bitstrings. The first constraint ensures that each bead is assigned to exactly one lattice site by penalizing more than one conformation per bead. The second constraint involves calculating a self-avoidance term that counts the number of beads that have chosen the same conformation, thus preventing two amino acids from occupying the same lattice point. The third constraint maintains sequence order by checking the distance between consecutive beads on the string~\cite{irback_folding_2022}.

\paragraph*{Turn-based HP-lattice model.}
We describe the protein conformation by tracking the directions in which each amino acid turns when placed sequentially on the lattice. The distance between beads and interactions in the HP-lattice energy function can be rewritten accordingly~\cite{babbush_construction_2014, babej_coarse-grained_2018, robert_resource-efficient_2021}.

This turn-based encoding of the beads' placements reduces the possible values of the location variables at the cost of increasing the number of interactions. The elements of the cardinality vector defined in Eq.~(\ref{eq:card_vector}) correspond here to the lattice coordination number, and thus
\begin{align}
    \cardvec_{\text{HP-square}}^{\text{turn}} = \ & (4, \dots,4),\\
    \cardvec_{\text{HP-cubic}}^{\text{turn}} = \ & (6, \dots,6).
    \label{eq:card_turn}
\end{align}

The simplicity of encoding the conformation for each bead in the sequence makes for a complex energy function that calls for extra resources, i.e., additional qubits.
Euclidean distances between beads are computed by summing the turns in all directions and converting them into coordinates based on the protein's conformation. Auxiliary qubits are employed to store these distances, enabling the calculation of pairwise interactions between sequence beads. 
The auxiliary bits can also be used to ensure that the distance between all beads is non-zero, i.e., no overlapping beads.

There are two versions of the turn-based model present in the literature: the turn ancilla encoding, which places the interaction information in auxiliary qubits, and the turn circuit encoding, which uses multi-qubit terms to keep track of this information.
The first one leads to a quadratic qubit growth with the number of amino acids.
The latter uses non-unitary half-adders and XNOR-gates~\cite{fingerhuth_quantum_2018, babbush_construction_2014}, which may be a more resource-efficient encoding. We have not considered this encoding in our analysis, but note that many half-adders may add auxiliary bits to be implemented to the otherwise linear qubit requirement~\cite{barbosa_quantum_2006}.

\subsection{Side-chain conformation-based models}
\label{subsec:Conf_model}

Several protein folding algorithms alternate between side-chain packing and backbone optimization to predict the optimal protein fold. The side-chain conformation-based models are used to find the optimal side-chain packing given a fixed backbone sequence and variable side-chain conformations. In contrast to the HP-lattice model, these models can reach a higher resolution for this intermediate step as we carry more information on the amino acid composition.
We focus on the side-packing problem, which is NP-complete~\cite{pierce_protein_2002}. We consider a discrete number of torsional angles or rotamers describing a given backbone's amino acid side-chain conformations.
Therefore, following Eq.~(\ref{eq:card_vector}), for each amino acid $a_i$ on a sequence, its side chain can take a limited number of rotational conformations $c_i$, with $\prod_i \cardelement_i$ the number of all feasible protein structures.

In these rotamer-based models, a function determines the energy of each protein conformation and guides the optimization toward the most energetically favorable structures. A commonly used function is the Rosetta energy function, which combines physics-based potentials with knowledge-based energies to return an energy score for a given structure~\cite{alford_rosetta_2017}. This energy function is a fundamental component of the Rosetta software suite used for protein structure prediction and design, already considered for optimization with quantum annealers~\cite{mulligan_designing_2020}.

Without loss of generality, we use an energy-scoring function that considers all pairwise interactions between rotamers at different amino acids. That is, given the conformations $r_i \in R_i$ from the set of possible conformations $R_i$ of the side chain associated with each amino acid $a_i$, the energy function is given by
\begin{equation}
    E(\mathbf{r}) = \sum_{i=2}^N \sum_{j<i}^{N-1} E(r_i,r_j),
\end{equation}
with $\mathbf{r}=(r_1,r_2,\dots,r_N)$ a vector of integers associated with a particular rotamer selection and protein structure.

\section{Encodings of discrete variables into qubits}
\label{sec:DVtoqubits}
A function scoring the energy of a protein conformation $x=r_1 r_2 \dots r_N$ can be translated to a quantum cost Hamiltonian. With that aim, we consider the concatenation of every integer $r_i$ represented with binary variables, such that $x=x_1 x_2 \dots x_M$, with $x_i \in \{0,1\}$. Then, each classical variable $x_i$ maps to a quantum variable $\sigma_i^z$ that can take values $\{1,-1\}$ as $x_i \rightarrow (1-\sigma_i^z)/2$.
This cost Hamiltonian, used in quantum annealing~\cite{farhi_quantum_2000} and in gate-based quantum algorithms~\cite{farhi_quantum_2014}, can be represented as a $k$-local spin Hamiltonian
\begin{align}
\label{eq:Hcost}
    H_{\text{cost}} = \sum_i h_i \sigma_i^z + \sum_{ij} J_{ij} \sigma_i^z \sigma_j^z
    + \sum_{ijk} J_{ijk} \sigma_i^z \sigma_j^z \sigma_k^z + \dots,
\end{align}
with interaction terms involving at most $k$ qubits. Here, $h_i$, $J_{ij}$, and $J_{ijk}$ correspond to the single-qubit, two-qubit, and many-qubit energy coefficients, respectively. The Pauli $z$-matrix acting on the $j$th qubit is $\sigma_j^z$.
Analyzing the necessary resources involves focusing on this cost Hamiltonian, which encodes the problem and guides the system from the initial to the optimal final state. 

\subsection{\label{subsec:encodings} Encodings: \oh{}, \binary{} and \bubin{}}
As mentioned earlier, the protein conformation is encoded into a bitstring $x= r_1r_2\dots r_N$. This encoding involves selecting one conformation $r_i \in R_i$ for each amino acid.
Different encodings of the integers $r_i$ result in varying resource usage.

\paragraph*{\oh{}.}
In the \oh{} encoding, an integer is denoted by a solitary one in the binary representation, positioned at the bit corresponding to the integer's value. Each set of conformation for a given amino acid $a_i$ is encoded by a sub-string of length $\cardelement_i$, where a single 1 represents one conformation, and the rest of the sub-string is 0, see Table~\ref{tab:encodings}. Each substring of length $\cardelement_i$ has the Hamming weight one to ensure only one conformation is chosen. The total length of the bitstring is $\sum_i \cardelement_i$, see Table~\ref{tab:resources_coord_conf}, and the total Hamming weight is equal to the number of amino acids $M$.

\paragraph*{\binary{}.}
The \binary{} encoding represents an integer by the binary numeral system where each digit's place value is a power of two, starting from the rightmost digit. Each set of conformations for a given amino acid $a_i$ is encoded by a sub-string of binaries, using $n$ bits so that $2^n\leq \cardelement_i$, see Table~\ref{tab:encodings}. We can not use the Hamming weight to check if a bitstring is in $F$. The total length of the bitstring is $\sum_i\ceil*{ \log_2 \cardelement_i }$, see Table~\ref{tab:resources_coord_conf}.

\paragraph*{\bubin{}.}
Block-Unary (BU) encodings can be considered as a balance between the \oh{} and \binary{} encodings, being more tunable to the limitations of the hardware. Block Unary contains blocks of size $g$, each with a \binary{} encoding, see Table~\ref{tab:encodings}. Each block is an element in a \oh{} string, where each block can encode for $g$ variables, and the rest of the bits in the other blocks are set to zero. The all-zero state denotes when the block is inactive. This allows each block to encode $2^n - 1 = g$, where $n$ represents the number of bits in a block. The total length of the bitstring is $\sum_i\ceil*{\frac{\cardelement_i}{g}}\ceil*{\log_2(g+1)}$, see Table~\ref{tab:resources_coord_conf}.

\begin{table}[htbp]
    \caption{\label{tab:encodings} The integer to binary \oh{}, \binary{}, and Block Unary Binary encodings with block size variable $g=3$.} 
\begin{ruledtabular}
\begin{tabular}{LLLL}
            \text{Decimal} & \text{\oh} & \text{\binary} & \text{\bubin}_{g=3} \\ \hline 
             0 & 10000 & 000 & 00 \ 01\\
             1 & 01000 & 001 & 00 \ 10\\
             2 & 00100 & 010 & 00 \ 11\\
             3 & 00010 & 011 & 01 \ 00\\
             4 & 00001 & 100 & 10 \ 00\\
        \end{tabular}
\end{ruledtabular}
\end{table}

\subsection{\label{subsec:compiling}Compiling to different quantum hardware}
To implement a parameterized quantum operation related to the cost Hamiltonian in a quantum circuit, we represent it as $e^{i\gamma H_{\text{cost}}}$ the with real parameter $\gamma$, mapping the quantum variables to logical qubits. This mapping relies on the specific operations supported by the target quantum processor, called the native gate set. The conversion requires breaking down complex operations in the cost Hamiltonian into elementary gates from the native gate set.
Different quantum hardware architectures vary in both connectivity and the available number of qubits. Platforms may be able to operate with many-qubit interactions but offer a lower number of qubits, e.g., the ion-trapped architecture~\cite{bruzewicz_trapped-ion_2019, muller_simulating_2011}, compared to platforms offering a higher number of qubits that may be constrained to two-qubit interactions, e.g., the superconducting circuit architecture~\cite{kjaergaard_superconducting_2020}.

In the latter case, one must decompose the many-qubit interactions into less connected gates. Compiling the multi-qubit interactions of the quantum algorithms analyzed here into CNOT and single-qubit gates gives rise to a quadratic overhead; see Appendix~\ref{app:compilation} for a detailed description. Other native gate sets and the availability of higher-order gates such as three-qubit gates~\cite{warren_extensive_2023} may reduce the depth of the quantum algorithms and required resources.
In Section~\ref{subsec:gate-k-loc}, we present how many $k$-qubit operations appear in the cost Hamiltonians of the protein folding models and how many one- and two-qubit operations would be needed to decompose the higher order operations given a limited native set.

During the compilation process, optimization techniques can be applied to improve the overall efficiency and performance of the quantum circuit. These techniques include gate merging, gate cancellation, and gate reordering to minimize the circuit depth. 
We assume all-to-all connectivity and do not consider the distance between qubits, which could lead to decoherence. However, if the hardware lacks all-to-all connectivity, a naive qubit routing procedure might demand an additional circuit depth of $\mathcal{O}{(n^3)}$. To tackle this problem, a SWAP network can be employed, offering a quadratic reduction in circuit depth compared to naive routing, resulting in a linear increase in operations~\cite{kivlichan_quantum_2018, ogorman_generalized_2019, hashim_optimized_2022}.

\section{\label{sec:resource_trade} Resource trade-offs}

We generate instances of folding and packing problems with amino acid chain lengths $N$ ranging from three to a hundred amino acids for five models, described in Sec.~\ref{sec:Coarse-grained protein models}, to compare the quantum resources required. For each problem instance, we have calculated the resources needed to form the variational state for the corresponding cost Hamiltonian, depending on the encoding used: \oh{}, \binary{}, or \bubin{}.

\begin{figure}[htbp]
    \includegraphics[width=0.44\textwidth]{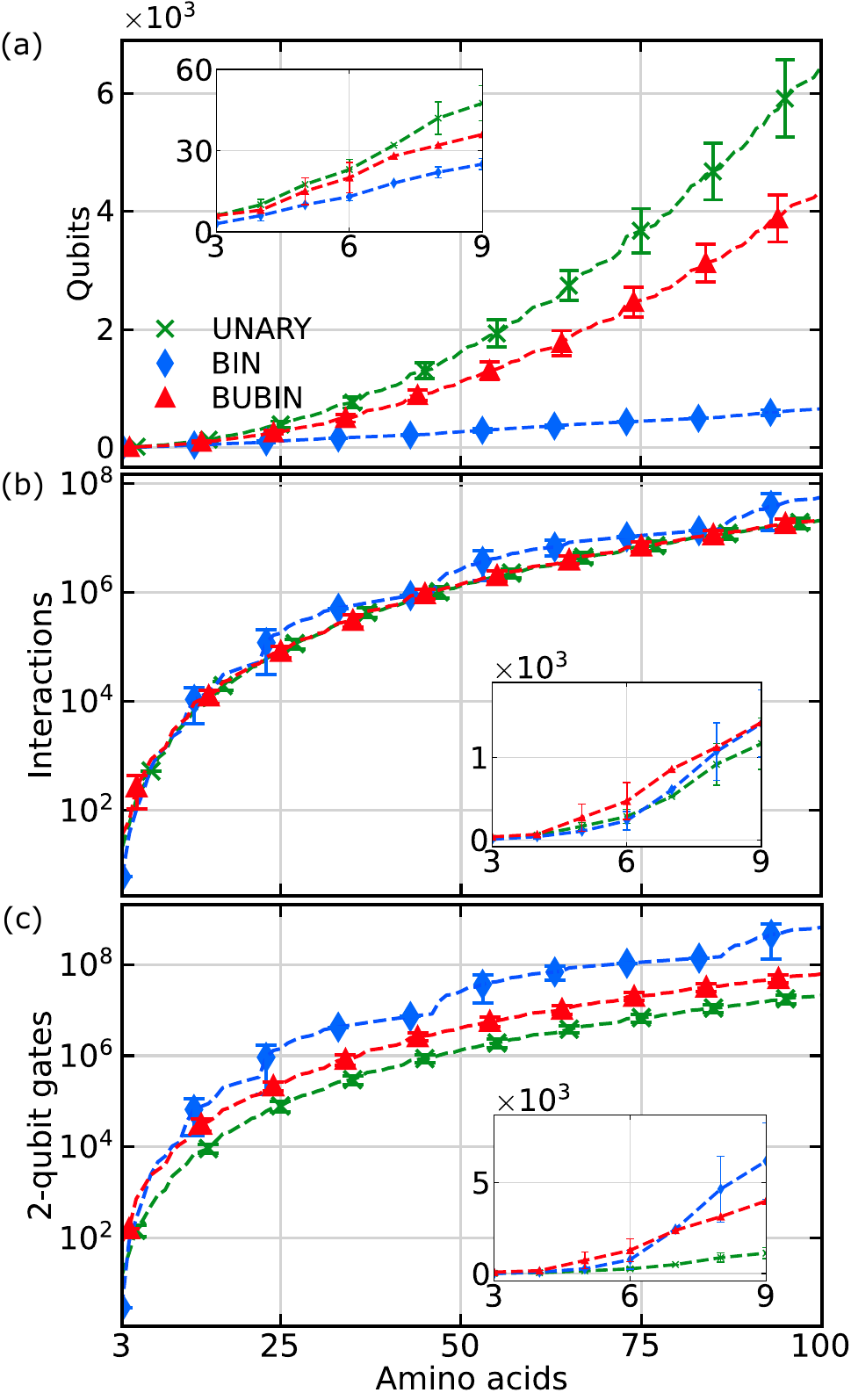}
    \caption{\textbf{Coordinate-based model on the square lattice.} The number of (a) qubits, (b) interactions, and (c) two-qubit gates required to implement a parametrized quantum operation based on the cost Hamiltonian: $e^{-i\gamma H_{\text{cost}}}$, with $H_{\text{cost}}$ given in Eq.~(\ref{eq:Hcost}), and a real parameter $\gamma$.
    We plot the required resources with \oh{} (green cross), \binary{} (blue rhombus), and \bubin{} (red triangle) encodings as a function of the number of amino acids $N$. The problem instance size ranges from $N=3, \dots,100$, and each figure inset zooms in on the results for fewer amino acids $N=3,\dots, 9$. For each $N$, we consider all rectangular grids with an area (number of sites) ranging from the lower bound $N$ to the upper bound 50\% larger than the lower bound, $1.5N$.
    The bars indicate one standard deviation.    
    }
    \label{fig:2dcoord}
\end{figure}

\begin{figure}[htbp]
    \includegraphics[width=0.44\textwidth]{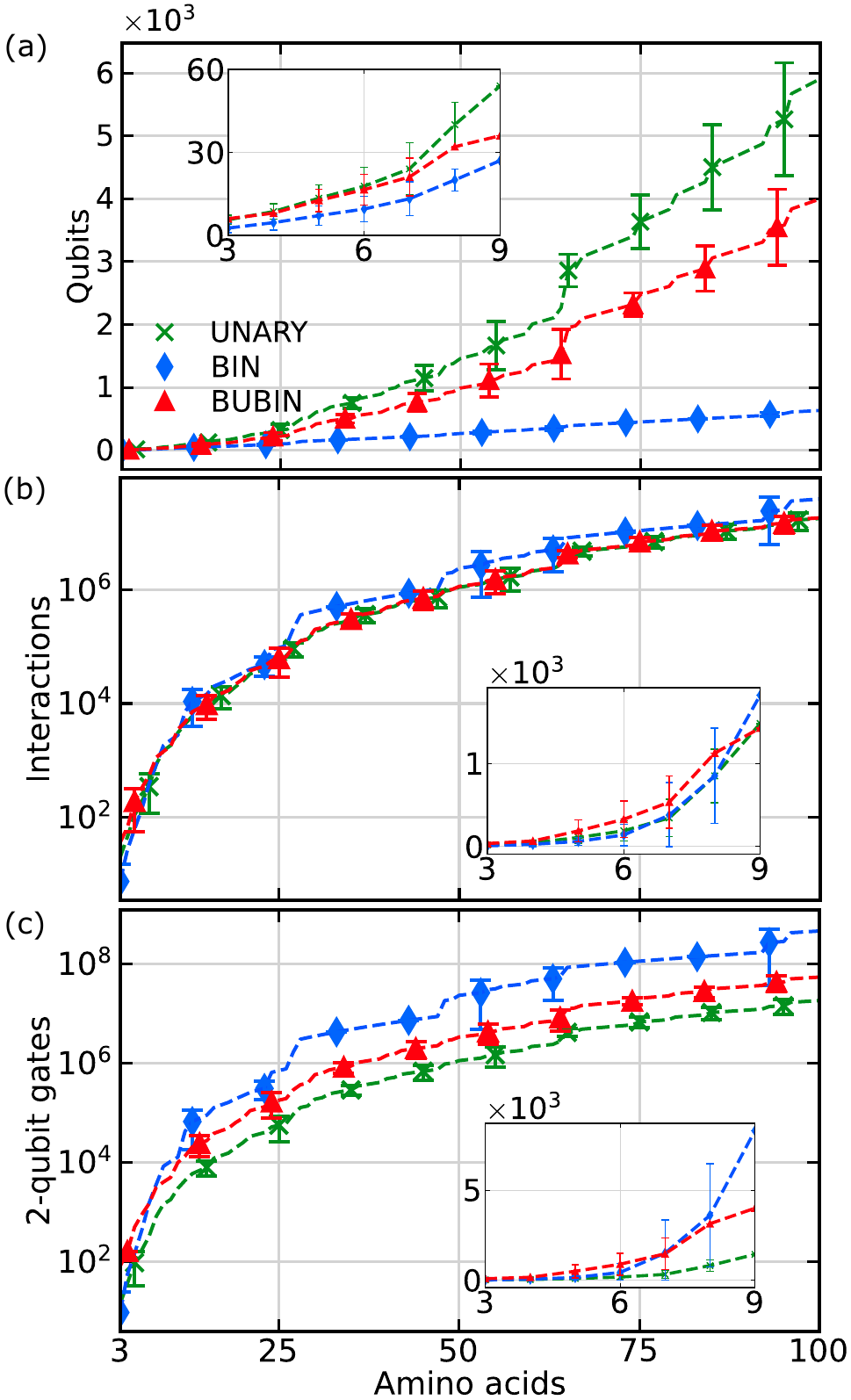}
    \caption{\textbf{Coordinate-based model on the cubic lattice.} The number of (a) qubits, (b) interactions, and (c) two-qubit gates required to implement a parametrized quantum operation based on the cost Hamiltonian: $e^{-i\gamma H_{\text{cost}}}$, with $H_{\text{cost}}$ given in Eq.~(\ref{eq:Hcost}), and a real parameter $\gamma$.
    We plot the required resources with \oh{} (green cross), \binary{} (blue rhombus), and \bubin{} (red triangle) encodings as a function of the number of amino acids $N$. The problem instance size ranges from $N=3, \dots,100$, and each figure inset zooms in on the results for fewer amino acids $N=3,\dots, 9$. For each $N$, we consider all cubic grids with a volume (number of sites) ranging from the lower bound $N$ to the upper bound 50\% larger than the lower bound, $1.5N$.
    The bars indicate one standard deviation.}
    \label{fig:3dcoord}
\end{figure}

Figures~\ref{fig:2dcoord} and \ref{fig:3dcoord} present the resources for the coordinate-based HP model on the two- and three-dimensional lattice, respectively.
The number of resources depends on the size of the lattice one chooses to fold the amino acid sequence on.
We calculate resources for all rectangular and cubic grids containing a total number of lattice sites between $N$ and $1.5N$. This range yields approximately ten problem instances for each sequence length $N$. Therefore, the presented resources represent averages and include one standard deviation.
The resources for turn-based lattice models directly depend on the number of amino acids $N$.
Figures~\ref{fig:2dturn} and \ref{fig:3dturn} illustrate the number of resources needed for encoding the two- and three-dimensional cases, respectively.
The side-chain conformation-based model's resources depend on the number of considered choices for each side-chain. We generate 2000 instances to account for different proteins. For each sequence length $N$, the number of conformations for each amino acid is drawn from a uniform distribution between two and a hundred. The average and standard deviation for the corresponding resources are shown in Fig.~\ref{fig:rosetta_all}. 

Furthermore, we compare our resource results to Ref.~\cite{robert_resource-efficient_2021}, where they consider a turn-based model on the tetrahedral lattice. The paper details the exact resource requirements of the number of qubits and interactions for up to 15 amino acids with fitted curves and presents the scaling of the number of terms in the Hamiltonian. Thus, we focus our comparison on the provided amino acid range of 15 and the scaling provided in the paper. Notably, we contrast the resource requirements of the tetrahedral lattice with the square lattice, due to their identical coordination numbers, despite their inherent differences.

In Sec.~\ref{subsec:qubit-growth}, the analysis focuses on the number of qubits required for representing the protein structure as a bitstring for the different models given the three encodings.
Section~\ref{subsec:gate-k-loc} studies the gates and many-body interactions needed, examining the computational complexity of implementing the various encoding schemes.
Lastly, Sec.~\ref{subsec:size_unfeasible} investigates the size of the unfeasible solution set, examining the challenges of finding valid protein conformations within the given constraints.

\begin{table*}[htbp]
    \caption{\label{tab:resources_coord_conf} Comparison of resources for the \oh{}, \binary{}, and \bubin{} encoding based on the cardinality vector $\cardvec$ from Eq.~(\ref{eq:card_vector}). The functions are valid for the side-chain conformation-based and coordinate-based lattice models. Details about correction terms $\varepsilon_{\binary}$ and $\varepsilon_{\bubin}$ can be seen in Appendix~\ref{app:correction}.} 
\begin{ruledtabular}
\begin{tabular}{LLLL}
            \text{Resource} & \text{\oh}   & \text{\binary} & \text{\bubin} \\ \hline
             \text{Qubits} & \sum_i \cardelement_i & \sum_i\ceil*{ \log_2 \cardelement_i } & \sum_i\ceil*{\frac{\cardelement_i}{g}}\ceil*{\log_2(g+1)}\\
             k\text{-Locality}  &  2 &  \ceil*{ \log_2(\text{max}(\cardelement_i))} + \ceil*{ \log_2(\text{max}_2(\cardelement_i))} & 2\ceil*{\log_2(g+1)}\\
             \text{Interactions}     & \sum_i \cardelement_i + \binom{\sum_i \cardelement_i}{2}       & \sum_i\ceil*{ \log_2 \cardelement_i} + \binom{\sum_i\ceil*{ \log_2 \cardelement_i}}{2} + & \sum_i\ceil*{\frac{\cardelement_i}{g}}\ceil*{\log_2(g+1)} + \binom{\sum_i\ceil*{\frac{\cardelement_i}{g}}\ceil*{\log_2(g+1)}}{2} +\\
& & \sum_{i<j}\sum_{m=3}^k\binom{\ceil*{ \log_2 \cardelement_i }+\ceil*{ \log_2 \cardelement_j}}{m} - \varepsilon_{\binary} & \binom{\sum_i\ceil*{\frac{\cardelement_i}{g}}}{2}\sum_{m=3}^{2\ceil*{\log_2(g+1)}}\binom{2\ceil*{\log_2(g+1)}}{m} - \varepsilon_{\bubin}\\
        \end{tabular}
\end{ruledtabular}
\end{table*}

Here, we summarize our main findings based on the analysis.
Firstly, our calculations demonstrate that the minimum number of required qubits needed for encoding a protein with a hundred amino acids in any of the considered models is around 600. This baseline is drawn by the \binary{} encoding in the side-chain conformation-based and coordinate-based lattice model.
Secondly, the \oh{} encoding requires the highest number of qubits, but the \binary{} encoding needs the most interactions and has a higher locality. The size of the unfeasible solution set---bitstrings corresponding to non-valid solutions---is smaller for the \binary{} encoding. The \bubin{} encoding strikes a middle ground between the other two encodings. The block size variable $g$ only shifts the \bubin{} encoding closer to or further from the \binary{} encoding with a larger or smaller $g$, respectively. For the sake of simplicity, we have used $g=3$ throughout the manuscript.
Thirdly, the difference between folding on a two- or three-dimensional lattice is negligible compared to the total number of resources for sequence length $N$, both in the number of qubits and interactions.
Finally, we note that the turn-based model on the tetrahedral lattice with the turn-based modelization of Ref.~\cite{robert_resource-efficient_2021} is generally more resource-efficient. Still, there are lattice sizes where the coordinate-based model with the \binary{} encoding uses the least amount of qubits and interactions to represent the same number of amino acids.
Notably, the coordinate-based model has a higher $k$-locality for the \binary{} encoding than the turn-based model on the tetrahedral lattice. The opposite is true for the \oh{} encoding, where the coordinate-based model has a lower $k$-locality.

In the insets of the figures, we present the results for reduced chain lengths $N=3,\dots, 9$, and for the side-chain conformation-based model, we reduce maximum number of side-chain conformations to twenty, averaging over 6000 instances for each sequence length $N$. Refer to Table~\ref{tab:resources_coord_conf} for an overview of the required resources calculated analytically based on the cardinality vector $\cardvec$ from Eq.~(\ref{eq:card_vector}) for the coordinate-based model and side-chain conformation-based model. For a detailed breakdown of the required number of terms and auxiliary qubits as a function of sequence length $N$ for the turn-based lattice model, please refer to Appendix~\ref{app:turnbased}.

\subsection{\label{subsec:qubit-growth} Number of qubits}
\begin{figure}[htbp]
    \includegraphics[width=0.44\textwidth]{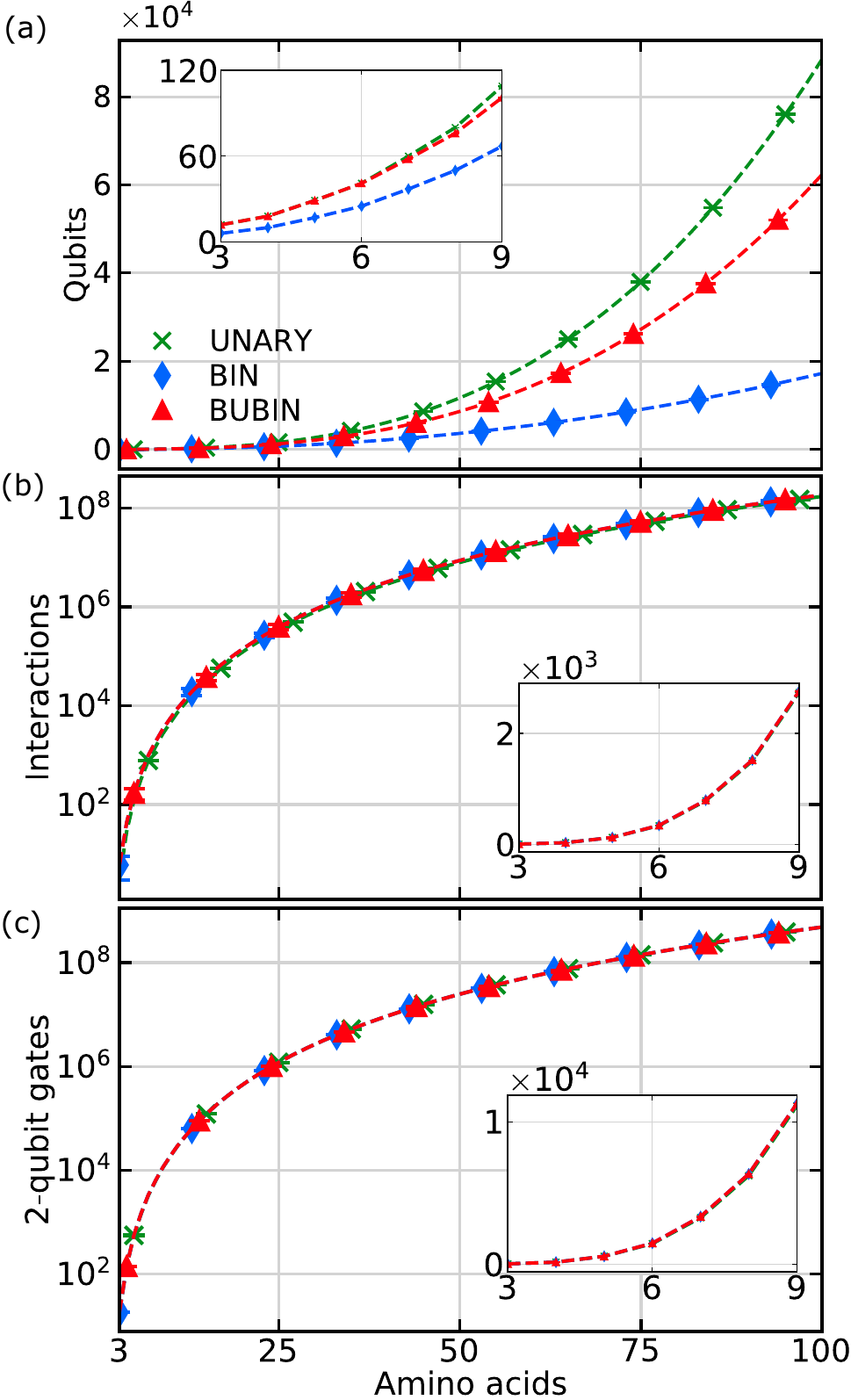}
    \caption{\textbf{Turn-based model on the square lattice.} The number of (a) qubits, (b) interactions, and (c) two-qubit gates required to implement a parametrized quantum operation based on the cost Hamiltonian: $e^{-i\gamma H_{\text{cost}}}$, with $H_{\text{cost}}$ given in Eq.~(\ref{eq:Hcost}), and a real parameter $\gamma$.
    We plot the required resources with \oh{} (green cross), \binary{} (blue rhombus), and \bubin{} (red triangle) encodings as a function of the number of amino acids $N$. The problem instance size ranges from $N=3, \dots,100$, and each figure inset zooms in on the results for fewer amino acids $N=3,\dots, 9$.}
    \label{fig:2dturn}
\end{figure}

\begin{figure}[htbp]
    \includegraphics[width=0.44\textwidth]{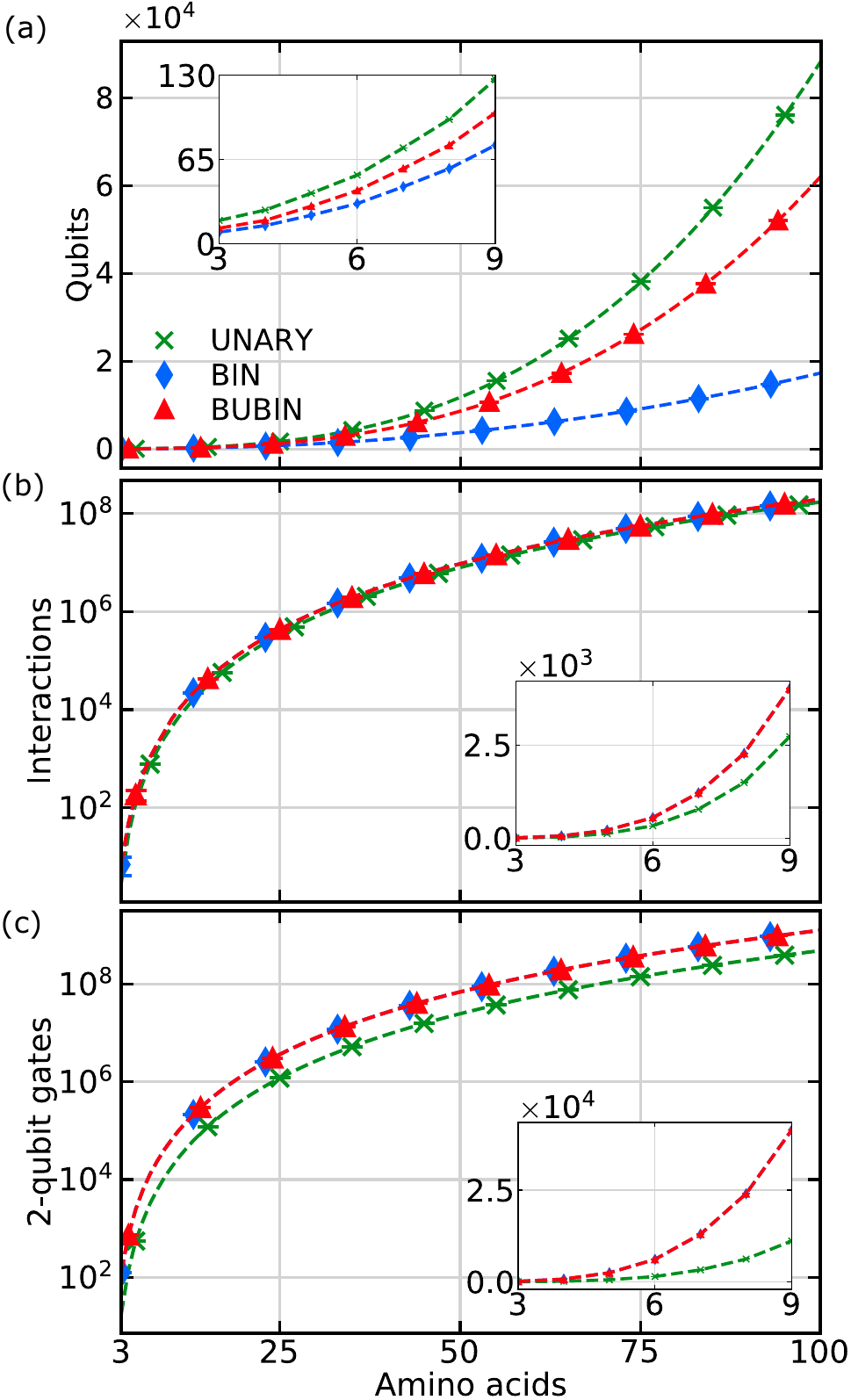}
    \caption{\textbf{Turn-based model on the cubic lattice.} The number of (a) qubits, (b) interactions, and (c) two-qubit gates required to implement a parametrized quantum operation based on the cost Hamiltonian: $e^{-i\gamma H_{\text{cost}}}$, with $H_{\text{cost}}$ given in Eq.~(\ref{eq:Hcost}), and a real parameter $\gamma$.
    We plot the required resources with \oh{} (green cross), \binary{} (blue rhombus), and \bubin{} (red triangle) encodings as a function of the number of amino acids $N$. The problem instance size ranges from $N=3, \dots,100$, and each figure inset zooms in on the results for fewer amino acids $N=3,\dots, 9$.}
    \label{fig:3dturn}
\end{figure}

\begin{figure}[tbp]
    \includegraphics[width=0.44\textwidth]{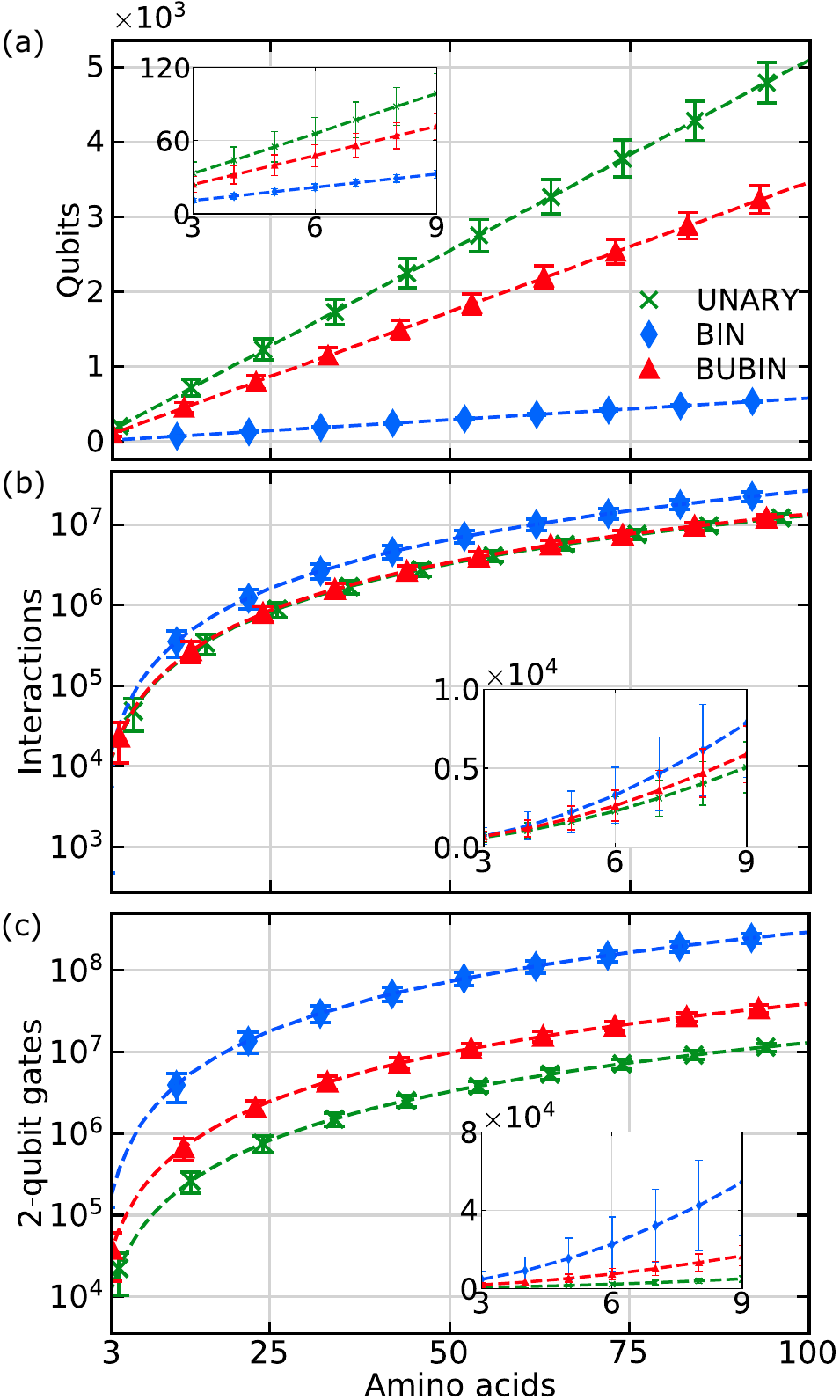}
    \caption{\textbf{Side-chain conformation-based model.}     The number of (a) qubits, (b) interactions, and (c) two-qubit gates required to implement a parametrized quantum operation based on the cost Hamiltonian: $e^{-i\gamma H_{\text{cost}}}$, with $H_{\text{cost}}$ given in Eq.~(\ref{eq:Hcost}), and a real parameter $\gamma$.
    We plot the required resources with \oh{} (green cross), \binary{} (blue rhombus), and \bubin{} (red triangle) encodings as a function of the number of amino acids $N$. The problem instance size ranges from $N=3, \dots, 100$, with the number of conformations for each amino acid uniformly distributed, $\mathcal{U}(2,100)$. Each figure inset zooms in on the results for fewer amino acids $N=3,\dots, 9$ and a reduced number of conformations, $\mathcal{U}(2,20)$. The bars indicate one standard deviation.}    
    \label{fig:rosetta_all}
\end{figure}

The subfigures (a) of Figs.~\ref{fig:2dcoord} to~\ref{fig:rosetta_all} show the number of qubits.
The required qubits grow linearly with the number of amino acids for the side-chain conformation-based models, as seen in Fig.~\ref{fig:rosetta_all} and Table~\ref{tab:resources_coord_conf}.
For the turn-based model, the number of required qubits is quadratic due to the required number of auxiliary qubits---needed to implement penalty terms mentioned in Sec.~\ref{subsec:HP_models}---see Fig.~\ref{fig:2dturn} and Appendix~\ref{app:turnbased}. 
The coordinate-based lattice model may look linear with the number of amino acids as observed in Table~\ref{tab:resources_coord_conf}, but as seen in Figs.~\ref{fig:2dcoord} and \ref{fig:3dcoord}, the number of needed qubits is quadratic. The quadratic behavior arises from the elements $\cardelement_i$ of the cardinality vector $\cardvec$ containing the lattice size, which, in turn, depends on the length of the amino acid sequence, as illustrated in Eq.~(\ref{eq:card_turn}).

The number of qubits for the \oh{} encoding for the coordinate-based model and the conformation-based is simply the sum of all elements in $\cardvec$.
The \binary{} encoding reduces the number of qubits needed, compared to the \oh{} encoding, to $\sum_i \ceil*{\log_2 \cardelement_i}$. 
The advantage of the \binary{} encoding becomes apparent in Figs.~\ref{fig:2dcoord}, \ref{fig:3dcoord}, and \ref{fig:rosetta_all}. It requires nearly an order of magnitude fewer resources than the \oh{} encoding. For the turn-based model, the improvement of using the \binary{} encoding is not as significant, see Figs.~\ref{fig:2dturn} and \ref{fig:3dturn}.
The \bubin{} encoding lies in between the \binary{} and \oh{} encoding, and the total number of qubits needed is the sum of all the blocks used times the number of qubits per block. The number of blocks is inversely proportional to $g$, $\sum_i \ceil*{\frac{\cardelement_i}{g} }$, and the number of qubits in a block is binary logarithmic in $g$, $\ceil*{\log_2(g+1)}$. As the all-zero-state in the blocks cannot be used to encode any information, each block encodes one less integer than the binary equivalent. As $g$ approaches the upper limit equal to the largest element of $\cardvec$, one obtains the \binary{} encoding without possibly using the all-zero-state. Using the lower limit $g=1$ yields the \oh{} encoding instead.

Comparing two- and three-dimensional models, the qubit difference is negligible relative to the total number of qubits required. For the coordinate-based model, the difference is only a few hundred because the minimum number of sites to allocate the amino acid chain is the same for two and three dimensions. The two-dimensional case has an average of more qubits as it is easier to pack the same number of sites tighter on a cubic lattice than on a square lattice.
For the turn-based model, the number of needed qubits is similar because the number of possible directions in three dimensions causes two extra qubits per bead, which is insignificant compared to the number of auxiliary qubits needed.

For a 15-amino acid chain encoded with \oh{}/\binary{} on the tetrahedral lattice, 76/53 qubits are needed. Compared to the other two models with coordination number four, the turn-based model requires 404/226 qubits, and the coordinate-based model needs an average of 135/52.5 qubits (SD = 15/7.5), with SD denoting standard deviation, for coordination number four.
The number of qubits required for the two models is similar, with the \binary{} encoding making the coordinate-based model more resource-efficient than the turn-based model on the tetrahedral lattice.
If the smallest lattice sizes are used in the coordinate-based model, it outperforms the turn-based model on the tetrahedral lattice from 13 amino acids onwards.
The scaling of the required number of qubits for the turn-based tetrahedral lattice~\cite{robert_resource-efficient_2021} is comparable to the coordinate-based model with a qubit requirement scaling as $\numaminoacids^2$.

\subsection{\label{subsec:gate-k-loc} Gates and many-body interactions}
The subfigures (b) of Figs.~\ref{fig:2dcoord} to~\ref{fig:rosetta_all} indicate the number of terms in the cost Hamiltonian. The $k$-locality of the Hamiltonian---specifying that the Hamiltonian terms act on at most $k$ qubits---will depend crucially on the encoding. For specific hardware, the terms need to be decomposed into native gates. The higher the $k$-locality of the terms, the more two-qubit gates are required. We have chosen native two-qubit CNOT-gates as an example, see Appendix~\ref{app:compilation}, which is, e.g., relevant for the superconducting circuit platforms. The subfigures (c) show the number of two-qubit gates. Table~\ref{tab:resources_coord_conf} presents the analytical expressions for the number of gates and $k$-locality for the coordinate-based lattice models and the side-chain conformation model. For details about the number of gates required for the turn-based lattice model, see Appendix~\ref{app:turnbased}.
The \oh{} encoding in the coordinate-based and side-chain conformation-based models involves only one- and two-body interactions. Thereby, the curve for the \oh{} encoding remains the same in subfigures (b) and (c) in Figs.~\ref{fig:2dcoord}, \ref{fig:3dcoord} and \ref{fig:rosetta_all}. The number of one-body interactions with the \oh{} encoding is $\sum_i \cardelement_i$, with $\cardelement_i$ the number of possible conformations for the $i$th amino acid. The number of two-body interactions is then given by $\binom{\sum_i \cardelement_i}{2}$. There are no higher-order interactions, that is, the Hamiltonian is $2$-local.

The \binary{} encoding will have a higher $k$ in the $k$-locality than the \oh{} encoding. For the coordinate-based and side-chain conformation-based models, the \binary{} encoding $k$-locality depends on the two largest elements $\cardelement_i$ of the cardinality vector $\cardvec$. Each amino acid $a_i$ will need $\ceil{\log_2{\cardelement_i}}$ qubits, and the two largest such values added together will yield the value of $k$. On average, the number of gates needed for the \binary{} encoding is much larger than \oh{}.
To pairwise connect amino acids $a_i$ and $a_j$, all the qubits associated with the two amino acids, i.e., $\ceil*{\log_2 \cardelement_i} + \ceil*{\log_2 \cardelement_j}$ need to be connected, resulting in a $m$-body ($m\geq3$) interaction requirement of $\binom{\ceil*{\log_2 \cardelement_i} + \ceil*{\log_2 \cardelement_j}}{m}$ for each pair. As explained in Sec.~\ref{sec:DVtoqubits}, each classical binary variable $x_i$ maps to $(1-\pauli{i}{z})/2$, and the multiplication of these binary variables leads to multi-qubit terms of all orders. The total number of $m$-body interactions needed is then given by summing over all $m$ and all pairs of amino acids. If $\ceil*{\log_2 \cardelement_i}\geq3$, we count higher-order interactions that are not unique to the pair and are present in every pair that includes $a_i$. These interactions could be grouped together and subtracted with a correction term $\varepsilon_\text{\binary}$, as seen in Appendix~\ref{app:correction}.
The number of one-body interactions coincides with the number of qubits $\sum_i \ceil*{ \log_2 \cardelement_i}$. The number of two-body interactions is $\binom{\sum_i \ceil*{ \log_2 \cardelement_i }}{2}$, that is, all combinations of two qubits.
The highest contribution to the number of gates for the \binary{} encoding will be the double sum over all amino acids and all higher order interactions $m=\{3,\dots, k\}$.

In the \bubin{} encoding, the $k$-locality of the Hamiltonian is set with the integer variable $g$, which means that for a given hardware with $k$-body gates, it is possible to set $g$ to match the hardware.
Similar to the \binary{} encoding, the number of one-body interactions in the \bubin{} encoding is the same as the number of qubits and is given by $\sum_i \ceil*{\frac{\cardelement_i}{g}}\ceil*{\log_2(g+1)}$.
The number of two-body interactions is equal to the number of pairwise combinations of all qubits $\binom{\sum_i \ceil*{\frac{\cardelement_i}{g}}\ceil*{\log_2(g+1)}}{2}$.
Analogously to the \binary{} encoding, the number of $m$-body interactions needed is calculated by considering pairwise interactions of blocks, as described in Sec.~\ref{subsec:encodings}, and summing over all $m$-body terms that contribute to a pair of blocks, written as $\binom{\sum_i\ceil*{\frac{\cardelement_i}{g}}}{2}\sum_{m=3}^{2\ceil*{\log_2(g+1)}}\binom{2\ceil*{\log_2(g+1)}}{m }$.
Additionally, if $g\geq4$, there will be interactions that appear multiple times and could be grouped, again we subtract a small correction term $\varepsilon_{\text{\bubin}}$, described in Appendix~\ref{app:correction}.

For the turn-based model, the $k$-locality of the interactions with the \oh{} encoding is also lower than that of the \binary{} encoding. However, in contrast to the other two models, this $k$-locality of the \oh{} encoding is now larger than two. The number of gates needed for the \oh{} is similar to, but still less than, the number of gates required for the \binary{} and \bubin{} encodings, see in Fig.~\ref{fig:2dturn} and \ref{fig:3dturn}.
Upon decomposition, the \oh{} encoding demands even fewer gates than the \binary{} encoding, especially evident in the three-dimensional case, as depicted in Fig.~\ref{fig:3dturn}.
Comparing Figs.~\ref{fig:2dcoord} and \ref{fig:3dcoord} to Figs.~\ref{fig:2dturn} and \ref{fig:3dturn}, it is evident that the coordinate-based model requires fewer interactions than the turn-based model in both the two- and three-dimensional cases.

When examining the number of interactions in the turn-based model on both square and tetrahedral lattices, in agreement with prior research~\cite{robert_resource-efficient_2021}, it is evident that the latter model uses fewer interactions and has a lower locality.
Our analysis shows that the coordinate-based model scales the same in the number of interactions as the turn-based model on the tetrahedral lattice. The coordinate-based has a scaling of $\mathcal{O}(N^4)$, see Table~\ref{tab:resources_coord_conf}, and the turn-based model on the tetrahedral lattice has $\mathcal{O}(\numaminoacids^4)$, as presented in Ref.~\cite{robert_resource-efficient_2021}. 
Comparing the interactions needed for encoding a 15 amino acid chain in the coordinate-based model calls for a higher number of gates in general. For the \oh{} encoding, the coordinate-based has 9292.5 terms on average (SD = 2032.5), and the turn-based model on the tetrahedral lattice has 4997 terms. For the \binary{} encoding, the coordinate-based model has 14550 terms on average (SD = 9300), and the turn-based model on the tetrahedral lattice has 9994 terms. Again, there are smaller lattices where the coordinate-based model using the \binary{} encoding calls for fewer resources.

\subsection{\label{subsec:size_unfeasible} Size of unfeasible solution set}

\begin{figure}[tbp]
    \centering
    \includegraphics[width=0.44\textwidth]{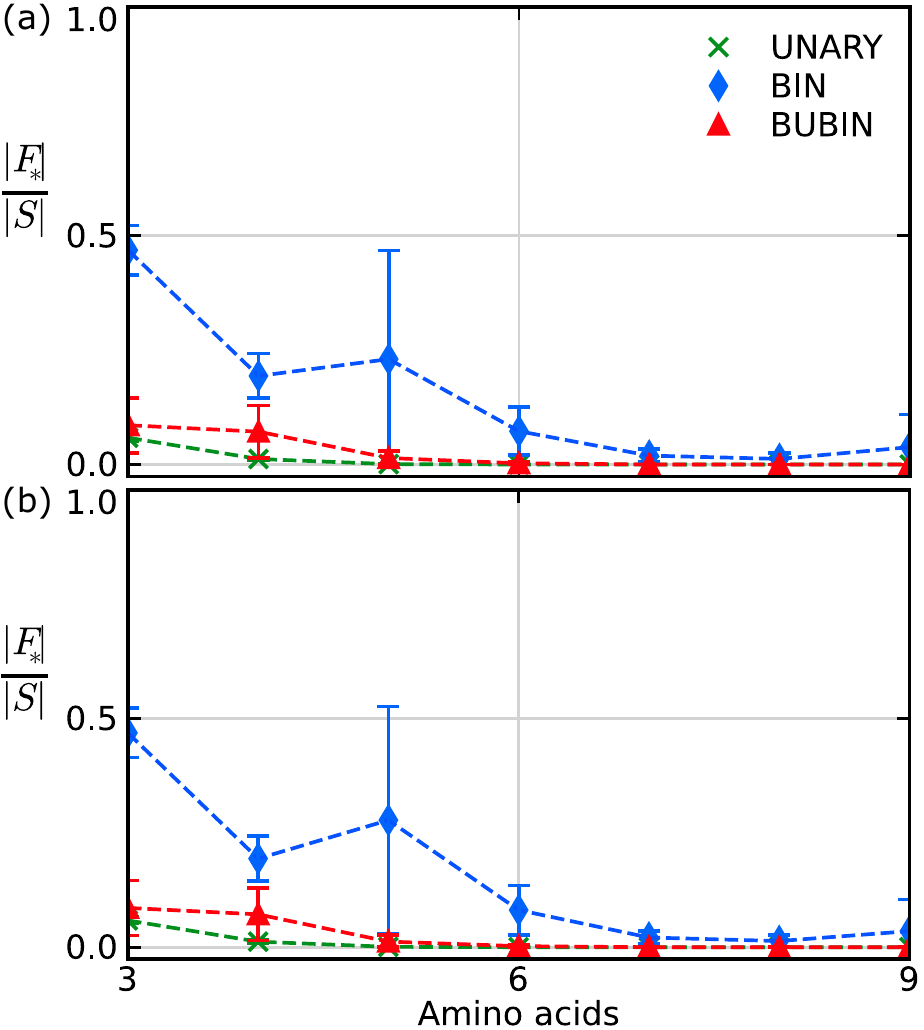}
    \caption{\textbf{Coordinate-based model on square (a) and cubic (b) lattices.} Ratio between the size of the approximate feasible solution set $|F_*|$ and the number of possible bitstrings $|S|$ as a function of the number of amino acids $N$ for the \oh{} (green cross), \binary{} (blue rhombus), and \bubin{} (red triangle) encodings. For each $N$, we consider all square/cubic grids with an area/volume (number of sites) ranging from the lower bound $N$ to the upper bound 50\% larger than the lower bound, $1.5N$.
    The bars indicate one standard deviation.}
    \label{fig:unfeasable_coord}
\end{figure}

The feasible solution set $F$ encompasses all bitstrings encoding valid physical solutions, where precisely one conformation is selected for each amino acid, and no constraints are violated.
The choice of the encoding for the problem will affect the size of this feasible solution set compared to all possible solution bitstrings.
Here, we discuss the ratio between the sizes of the feasible solution set $F$ and the total solution set $S$, which contains the feasible and unfeasible solution sets. We want the unfeasible solution set to be as small as possible so that it is easier to find the native structure of the protein.

\begin{figure}[htbp]
    \centering
    \includegraphics[width=0.44\textwidth]{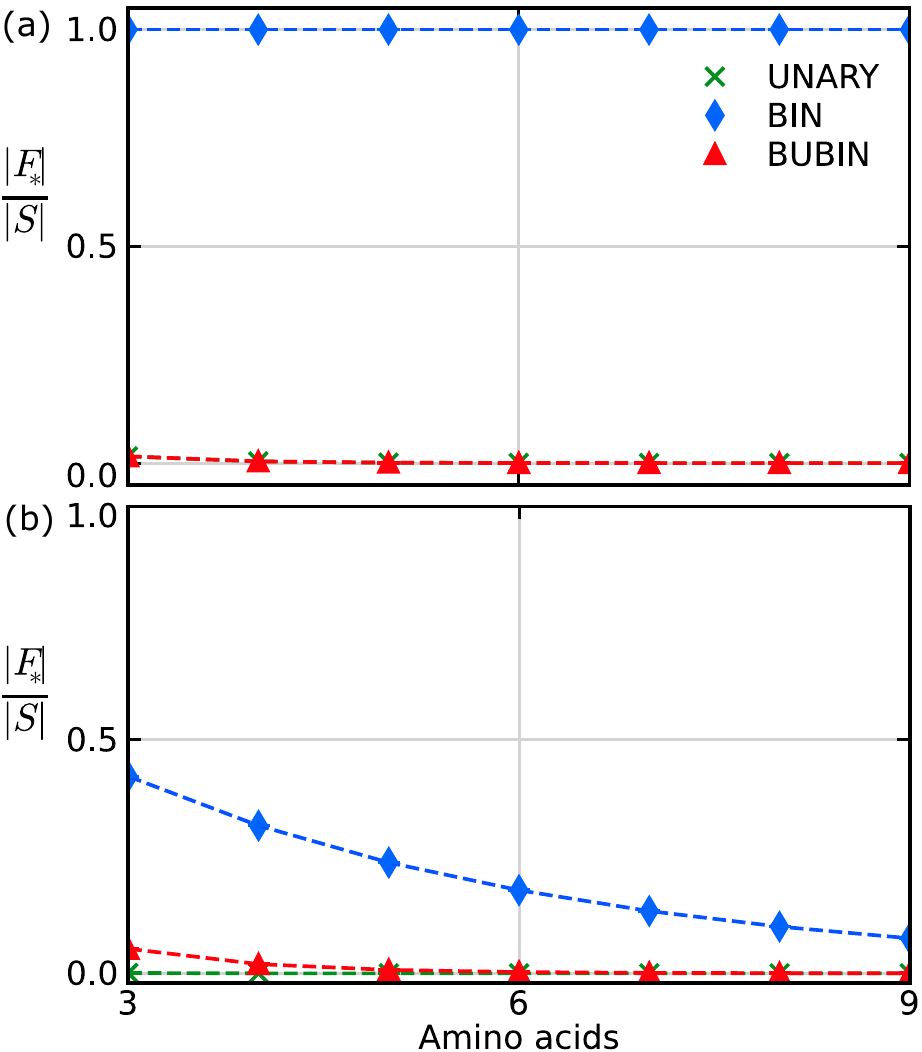}
    \caption{\textbf{Turn-based model on square (a) and cubic (b) lattices.} Ratio between the size of the approximate feasible solution set $|F_*|$ and the number of possible bitstrings $|S|$ as a function of the number of amino acids for the \oh{} (green cross), \binary{} (blue rhombus), and \bubin{} (red triangle) encodings.
    }
    \label{fig:unfeasable_turn}
\end{figure}

\begin{figure}[htbp]
    \centering
    \includegraphics[width=0.44\textwidth]{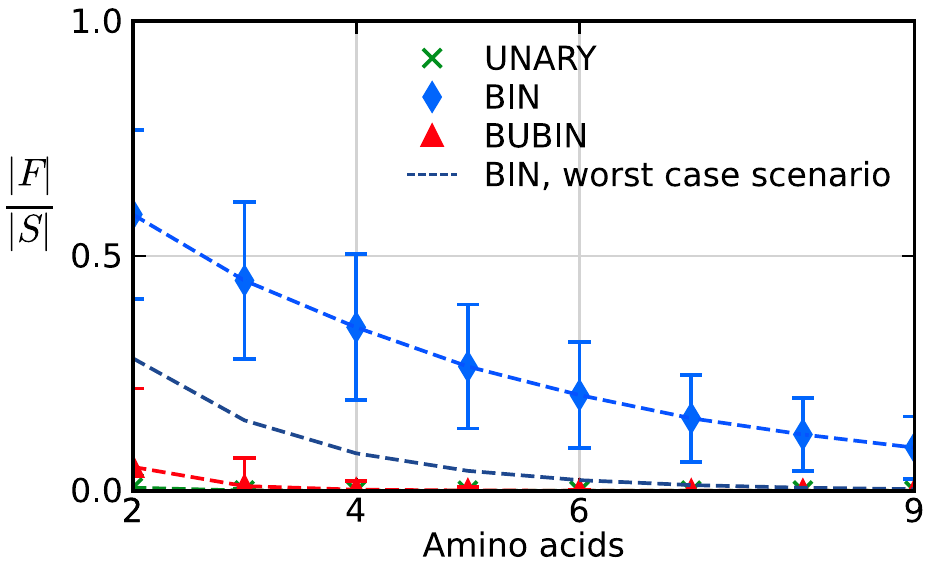}
    \caption{\textbf{Side-chain conformation-based models.} Ratio between the size of the feasible solution set $|F|$ and the number of possible bitstrings $|S|$ as a function of the number of amino acids $N$ for the \oh{} (green cross), \binary{} (blue rhombus), and \bubin{} (red triangle) encodings.
    The problem instance size ranges from $N=3, \dots, 9$, with the number of conformations for each amino acid uniformly distributed, $\mathcal{U}(2,20)$. The bars indicate one standard deviation.
    The dashed blue line without a marker is the worst-case scenario for the \binary{} encoding.}
    \label{fig:unfeasable_conf}
\end{figure}

The bitstrings constituting the feasible solutions for the coordinate-based lattice model are determined by three penalty terms presented in Sec.~\ref{subsec:HP_models}.
In Fig.~\ref{fig:unfeasable_coord}, we show an approximate upper bound of the relative size of the feasible solution set for the coordinate-based lattice model. We have only considered two model constraints: one amino acid per site and one site per amino acid. The approximate feasible solution set $F_*$ contains solutions that may have a broken chain.
Similarly, we present an approximate upper bound of $|F_*|/|S|$ for the turn-based model in Fig.~\ref{fig:unfeasable_turn}, as the calculations are only based on the cardinality vector in Eq.(\ref{eq:card_turn}) and do not contain the overlap constraint.
The auxiliary qubits of the turn-based lattice model are not accounted for in this section, as they do not affect the feasibility of the bitstrings.
For the side-chain conformation-based model, we present the exact results in Fig.~\ref{fig:unfeasable_conf}.
Table~\ref{tab:feasible_set} includes the analytical expression of the relative sizes of the feasible solution sets. The calculations are based on the bitstrings generated by choosing just one conformation per amino acid for the three encodings.

The relative size of the feasible solution set decreases rapidly with the increasing number of amino acids for all encodings. The turn-based model on the two-dimensional lattice stands out as the only exception, with a relative size exceeding 25\% from nine amino acids onwards, as shown in Fig.~\ref{fig:unfeasable_turn}. The \oh{} encoding will have a high number of unfeasible solutions, as we need to keep Hamming weight one for each block of $\cardelement_i$ qubits, with $\cardelement_i$ the elements of the vector in Eq.~(\ref{eq:card_vector}).
Thereby, for the \oh{} encoding, the ratio $|F|/|S|$ tends to zero as the number of conformations per amino acid increases. It is upper bounded by $2^{-\numaminoacids}$, where each amino acid only has two conformations.

In contrast, the \binary{} encoding will generally have a smaller unfeasible solution set than the \oh{} encoding, even if its size depends highly on the structure of the problem instance.
In the best-case scenario for the \binary{} encoding, the unfeasible solution set is empty when each amino acid has $2^l, l \in \mathbb{N}$ conformations.
The turn-based model in two dimensions is an example of a best-case scenario of the \binary{} encoding, with each amino acid having $2^2=4$ conformations, see Fig.~\ref{fig:unfeasable_turn}.
In the worst-case scenario for the \binary{} encoding, when each amino acid has $2^l+1$ conformations, the unfeasible solution set tends to zero as with the \oh{} representation. Both the worst-case and the general case of the \binary{} encoding are better than the \oh{} encoding, though, as seen in Fig.~\ref{fig:unfeasable_conf}.

Again, the \bubin{} encoding falls between the \oh{} and \binary{} encodings, but the relative size of the feasible set rapidly approaches zero, as seen in Figs.~\ref{fig:unfeasable_coord}, \ref{fig:unfeasable_turn}, and \ref{fig:unfeasable_conf}, since multiple blocks are required to encode conformations for each amino acid.

\begin{table}[htbp]
    \caption{\label{tab:feasible_set} Comparison of the relative size of the feasible solution set, $\frac{|F|}{|S|}$, for the \oh{}, \binary{}, and \bubin{} encoding. These results are solely based on choosing one conformation per amino acid and thereby exact for the side-chain conformation-based model and approximate for the two lattice models.} 
\begin{ruledtabular}
\begin{tabular}{LL}
            \text{Encoding} & \text{Relative size of feasible set} \\ \hline 
            \text{\oh{}} & \prod_i \cardelement_i / 2^{\cardelement_i} \\
            \text{\binary} & \prod_i \cardelement_i / 2^{\ceil*{\log_2(\cardelement_i)}} \\
            \text{\bubin} & \prod_i \cardelement_i / 2^{2\ceil*{\frac{\cardelement_i}{g}}} \\
        \end{tabular}
\end{ruledtabular}
\end{table}

\section{\label{sec:conclusion} Conclusion and outlook}

In summary, we have analyzed the resources for five coarse-grained models: HP models on the lattice, with a turn-based and coordinate-based representation of the amino acids locations, and the off-lattice side-chain packing model. In particular, we have computed the required qubits, interactions, and two-qubit gates for problem formulations associated with three encodings: \oh{}, \binary{}, and \bubin{}.

We conclude that current NISQ devices are unsuitable for simulating protein folding instances larger than a proof-of-concept due to the significant gate requirement. 
The number of gates needed to address average-sized human proteins reaches at least $10^7$, which is significantly greater than what is now possible with current gate fidelities of 99.9\%~\cite{kosen_building_2022,cai_entangling_2023}. 
However, the necessary number of qubits is attainable within the NISQ era~\cite{noauthor_ibm_2015}, with the side-chain packing problem and protein folding using coordinate-based lattice models being the most feasible applications.
For instance, when employing the \binary{} encoding, a chain of a hundred amino acids can be represented with fewer than a thousand qubits on average.

When comparing the resource requirements of the three encodings, we observe that using the \binary{} representation calls for the least number of qubits and yields a smaller unfeasible solution set, but translates to the highest-order interactions in the Hamiltonian. The \binary{} encoding is thereby more suitable for quantum hardware with access to multi-qubit gates, like ion-trap computers.
Conversely, hardware limitations for the $k$-locality of the quantum gates, such as with superconducting qubits, favor using the \oh{} encoding, even if it requires more qubits to represent the same problem instance. 
The \bubin{} encoding, with the flexibility of choosing the block size, can strike a balance between the \oh{} and \binary{} encodings to accommodate specific device limitations. Our conclusions echo earlier results analyzing encodings for quantum simulations of quantum models~\cite{sawaya_resource-efficient_2020}.

Counterintuitively, we find that the coordinate-based model is more resource-efficient than the turn-based model for the square and cubic lattice, both in time and space complexity.
Even if the turn-based model requires fewer qubits for encoding protein conformations, we need additional auxiliary qubits to prevent overlapping conformations and encode pairwise interactions. These additional qubits significantly increase the qubit requirement by one order of magnitude compared to the coordinate-based model.
Further, the improved turn-based model on the tetrahedral lattice~\cite{robert_resource-efficient_2021} scales as the coordinate-based model on the square lattices in the number of qubits and gates needed. For shorter chains with a maximum of 15 amino acids, the turn-based model on the tetrahedral lattice generally uses fewer resources.
Still, the coordinate-based model with the \oh{} encoding has a lower $k$-locality than the tetrahedral model, and with the \binary{} encoding, it requires fewer qubits and gates for small lattices.

Even with a resource-efficient model, using a hybrid quantum algorithm does not guarantee finding the native structure of the protein. As presented by previous work~\cite{fingerhuth_quantum_2018,saito_lattice_2023, boulebnane_peptide_2022}, the more challenging part of achieving a quantum advantage in protein folding simulations may be the classical optimization of the parameters in the quantum circuit.
These obstacles generate skepticism about QAOA's ability to address the protein folding problem in the near future~\cite{boulebnane_peptide_2022}. Research in optimizing the quantum circuit goes hand in hand with the search for resource-efficient model formulations. 
A potential approach to address the requirement for auxiliary qubits could involve applying unequal penalization to the inequality constraints~\cite{montanez-barrera_unbalanced_2023}.
We hope our study will spur exploration for more resource-efficient models for protein folding and stimulate further research into qualitatively better quantum computing systems.

\begin{acknowledgments}
We thank Anders Irbäck, Lucas Knuthson, Sandipan Mohanty, and Carsten Peterson for the essential discussions that made this work what it is. We thank Ivano Tavernelli for helpful discussions and for sharing details of his paper's results.
We acknowledge support from the Knut and Alice Wallenberg Foundation through the Wallenberg Center for Quantum Technology (WACQT). L.G.-{\'{A}.} further acknowledges support from the Swedish Foundation for Strategic Research (grant number FUS21-0063) and OpenSuperQ-Plus100 (101113946).
\end{acknowledgments}

\section{Code availability}

The code used in this work can be found here: https://github.com/HannaLinn/resources-qaoa-protein-folding.
All calculations are performed in Python with Numpy~\cite{harris_array_2020} and SciPy~\cite{2020SciPy-NMeth}, and all the plots are generated with Matplotlib~\cite{hunter_matplotlib_2007}.

\appendix

\section{Correction term for \binary{} and \bubin{} encoding}
\label{app:correction}
\input{app_correction_term}

\section{Compilation of multi-qubit gates into universal gate sets}
\label{app:compilation}
\input{app_compilation}

\section{Locality and interaction in the turn-based lattice model}
\label{app:turnbased}
\input{app_turnbased}

\bibliography{references}

\end{document}

%% file: app_correction_term.tex
In Sec.\ref{subsec:gate-k-loc} of the main text, we calculate the number of interactions in the cost Hamiltonian of the side-chain conformation-based and coordinate-based lattice models. 
To compute the number of higher-order Hamiltonian terms for the \binary{} encoding, we consider all pairwise amino acid interactions and all combinations of multi-qubit terms ranging from order three to the total number of qubits associated with each amino acid pair. We need all the terms because, as explained in Sec.~\ref{sec:DVtoqubits}, each classical binary variable $x_i$ maps to $(1-\pauli{i}{z})/2$, and the multiplication of these binary variables leads to multi-qubit terms of all orders.
An amino acid $a_i$ has $c_i$ possible conformations, as shown in Eq.~(\ref{eq:card_vector}). 
If the amino acid $a_i$ is encoded by at least three qubits, a binomial term accounting for all possible $m$-element qubit combinations (with $m\geq 3$) will include some terms that only involve qubits encoding the amino acid $a_i$. However, we only want to count these $m$-element terms once, and therefore, we need to subtract them for every pairwise interaction in which the amino acid $a_i$ is included except one.

We can group these extra multi-qubit interactions in correction terms and subtract them from the total number. The correction terms are given by
\begin{equation}
    \varepsilon_\text{\binary} = \sum_i^N\sum_{m=3}^{\ceil*{ \log_2(\cardelement_i)}}\binom{\ceil*{\log_2(\cardelement_i)}}{m}(N-2),
\end{equation}
for the \binary{} encoding, where we assume that $N>2$.

For the \bubin{} encoding, instead of amino acids, we consider all pair-wise block interactions. By an analogous argument, the correction term is given by
\begin{equation}
    \varepsilon_\text{\bubin} = \left(-2+\sum_i^N\ceil*{\frac{\cardelement_i}{g}}\right) \sum_{m=3}^{\ceil*{\log_2(g+1)}}\binom{\ceil*{\log_2(g+1)}}{m}.
\end{equation}
Note that for the number of qubits that encode one block to be at least three, we need $\ceil{\log_2(g+1)}\geq3$, or $g\geq 4$. In this manuscript we used $g=3$, so $\varepsilon_\text{BUBinary}=0$. In general, both correction terms are small compared to the number of interactions.

%% file: app_compilation.tex
\begin{figure}[tbp]
\centering
\begin{quantikz} [row sep={0.45cm,between origins}, column sep=0.2cm]
\lstick{$q_i$}        & \ctrl{1} & \qw            & \qw      & \qw      & \qw               & \qw      & \qw      & \qw             & \ctrl{1} & \qw \\
\lstick{$q_{i+1}$}    & \targ{}  & \ctrl{1}       & \qw      & \qw      & \qw               & \qw      & \qw      & \ctrl{1}        & \targ{}  & \qw \\
\lstick{\vdots}     &          & \vdots         &          &          &                   &          &          & \vdots          &          \\
\lstick{$q_{j-1}$}    & \qw      & \targ{}\vqw{-1}& \ctrl{1} & \qw      & \qw               & \qw      & \ctrl{1} & \targ{}\vqw{-1} & \qw     & \qw \\
\lstick{$q_j$}        & \qw      & \qw            & \targ{}  & \ctrl{2} & \qw               & \ctrl{2} & \targ{}  & \qw             & \qw      & \qw\\
\lstick{\vdots}     &          &                &          &          &                   &          &          &                 &          \\
\lstick{$q_k$}        & \qw      & \qw            & \targ{}  & \targ{}  & \gate{R_z(\alpha)}& \targ{}  & \targ{}  & \qw             & \qw      & \qw\\
\lstick{$q_{k+1}$}    & \qw      & \targ{}\vqw{1} & \ctrl{-1}& \qw      & \qw               & \qw      & \ctrl{-1}& \targ{}\vqw{1}  & \qw     & \qw \\
\lstick{\vdots}     &          & \vdots         &          &          &                   &          &          & \vdots          &          \\
\lstick{$q_{\ell-1}$} & \targ{}  & \ctrl{-1}      & \qw      & \qw      & \qw               & \qw      & \qw      & \ctrl{-1}       & \targ{}  & \qw \\
\lstick{$q_{\ell}$}   & \ctrl{-1}& \qw            & \qw      & \qw      & \qw               & \qw      & \qw      & \qw             & \ctrl{-1} & \qw
\end{quantikz}
\caption{A circuit based on CNOT ladders implementing the Pauli tensor evolution in Eq.~(\ref{eq:phase-gadget-unitary}) in terms of two-qubit interactions.}
\label{fig:doblez_cnot}
\end{figure}
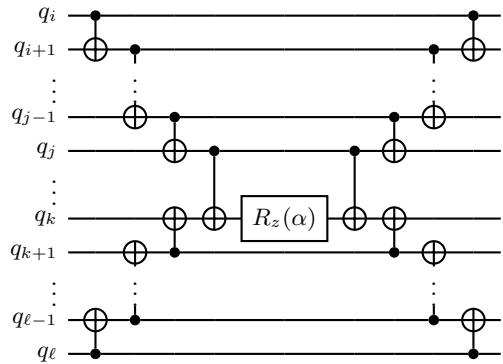

In Sec.~\ref{sec:resource_trade} of the main text, we calculate the number of two-qubit interactions in each of the five models, together with the encodings. To transition from $k$-body interaction to two-body interactions, we decompose the higher-order gates as follows.
The phase gadgets are a class of unitaries related to Pauli tensor evolution operators, essential blocks in variational quantum algorithms. 
The multi-qubit interactions relevant to the quantum algorithms analyzed in this work can be implemented in a superconducting circuit processor using CNOT-staircase constructions.
In particular, we consider the general operation
\begin{equation}\label{eq:phase-gadget-unitary}
    U^{ijk\ell}_{Z}(\alpha) = \exp\left[-i\frac{\alpha}{2}\pauli{z}{i} \cdots \pauli{z}{j}\pauli{z}{k} \cdots \pauli{z}{\ell})\right],
\end{equation}
which can be implemented with CNOT ladders, as shown in Fig.~\ref{fig:doblez_cnot}. With such construction, one needs $2(m-1)$ two-qubit gates for an $m$-body Pauli tensor, i.e., an interaction term acting upon $m$ qubits. Here, we do not consider reductions due to successive applications of these blocks or overheads for two-qubit gates between distant qubits.

This well-known CNOT-staircase construction is particularly suitable for quantum hardware with a gate set comprising single- and two-qubit operations, such as superconducting circuits. The final gate count depends on the particular available gate set and compilation. For instance, access to multiqubit gates~\cite{gu_fast_2021,molmer_multiparticle_1999} and controlled arbitrary-phase gates CZ$_\phi$ reduces the algorithm depth~\cite{lacroix_improving_2020}. Moreover, the particular qubit connectivity of the device introduces a gate overhead due to routing steps comprising additional SWAP gates that allow distant qubits to interact.

%% file: app_turnbased.tex
In Sec.~\ref{sec:resource_trade} of the main text, we calculate the number of qubits and interactions of the turn-based model on the square and cubic lattice.
Here, we review previous work on the turn-based lattice model. We pay attention to the locality of terms and count the interactions associated with each equation.
In Ref.~\cite{babbush_construction_2014}, the authors construct a cost Hamiltonian which encodes a two-dimensional ($D=2$) HP-lattice model with the \binary{} encoding.
In Ref.~\cite{babej_coarse-grained_2018}, the authors expand the earlier model to three dimensions ($D=3$) using the MJ energies, also with the \binary{} encoding.

\paragraph*{Cost Hamiltonian.}
The cost Hamiltonian has a general form consisting of three terms: one term that penalizes the choice of two consecutive turns folding back onto itself $H_{\text{back}}$, one term that penalizes overlapping later in the amino acid chain $H_{\text{overlap}}$, and $H_{\text{pair}}$ which is is the two-body interaction term (HP or MJ potential).

\paragraph*{Taking a turn.}
To begin, some equations are designed to represent turns in specific spatial directions. These equations evaluate as TRUE if the $j$-th turn occurs in the specified spatial direction, such as the $x$-direction:
\begin{align}
    d_{+x}^j &= (1-q_{3j-4}) q_{3j-5}q_{3j-3},\\
    d_{-x}^j &= (1-q_{3j-3}) q_{3j-5}q_{3j-4}.
\end{align}
These equations are then used to calculate how many times each turn has occurred in the previous chain, thereby getting the coordinate of the amino acid of interest.
In the \oh{} encoding, the turn equation for each direction only includes one qubit, as noted earlier~\cite{fingerhuth_quantum_2018}. Each equation has a locality of the lattice dimension $D$ in the \binary{} and \bubin{} case and one in the \oh{} case.

\paragraph*{Not turning back.}
The term $H_{back}$ uses AND functions ($\land$) that take two turns as input and return true if the second goes backward. For example, to penalize turning right, $+x$, at turn $j$ AND then left, $-x$, in turn $j+1$, we have the circuit
\begin{align}
    d_{+x}^j \land d_{-x}^{j+1} =&  [(1-q_{3j-4}) q_{3j-5}q_{3j-3}] \nonumber \\
     & [(1-q_{3j-3}) q_{3j-5}q_{3j-4}] \nonumber \\
     =& q_{3j-3}^2 q_{3j-4}^2 q_{3j-5}^2 - q_{3j-3}^2 q_{3j-4} q_{3j-5}^2 \nonumber \\
     &- q_{3j-3} q_{3j-4}^2 q_{3j-5}^2 + q_{3j-3} q_{3j-4} q_{3j-5}^2.
\end{align}
To penalize turning right then left OR ($\lor$) left then right, we have
\begin{align}
    (d_{+x}^{j} \land d_{-x}^{j+1}) \lor
    (d_{-x}^{j} \land d_{+x}^{j+1}) =& (d_{+x}^{j} \land d_{-x}^{j+1}) \nonumber \\
    &+ (d_{-x}^{j} \land d_{+x}^{j+1}) .
\end{align}
By combining all versions of right then left, left then right, and so on,
\begin{equation}
    (d_{+x}^{j} \land d_{-x}^{j+1}) \lor
    (d_{-x}^{j} \land d_{+x}^{j+1}) \lor
    (d_{+y}^{j} \land d_{-y}^{j+1}) \lor
    (d_{-y}^{j} \land d_{+y}^{j+1}),
\end{equation}
it is possible to penalize all backturns. The $H_{back}$ results in at most $2D$-local terms in the \binary{} and \bubin{} encodings, and $2$-local terms in the \oh{} encoding.
The final $H_{back}$ is then given by
\begin{align}
    H_{back} =& \lambda_{back} \Big\{ (q_0 \lor d_{-x}^2) + [(1-q_0)\lor d_{-y}^2] \nonumber \\
    +& \sum_{j=2}^{\numaminoacids-3} \big[ (d_{+x}^{j} \land d_{-x}^{j+1}) + (d_{-x}^{j} \land d_{+x}^{j+1}) \nonumber \\
    +& (d_{+y}^{j} \land d_{-y}^{j+1}) + (d_{-y}^{j} \land d_{+y}^{j+1}) \nonumber \\
    +& (d_{+z}^{j} \land d_{-z}^{j+1}) + (d_{-z}^{j} \land d_{+z}^{j+1}) \big] \Big\}.
\end{align}
The number of interactions on the three-dimensional lattice is $(\numaminoacids-5)$, from the summation, times the terms in the AND-expressions, which is $4\cdot2$ for the $x$-axis and $z$-axis and $12\cdot2$ for the $y$-axis and sums to twenty terms. The number of interactions on the two-dimensional lattice is $2\numaminoacids-10$, from the summation, times eleven.

\paragraph*{Position of the amino acids.}
Further, we need equations to describe the position of amino acid $i$ with the help of the boolean turn equations above, e.g., in the $x$-direction, we have:
\begin{equation}
    x_i = 
    \begin{cases}
      0, & \text{if }i=0 \\
      & \text{($a_0$ occupy the origin)},\\
      1 + q_0 +& \\
      \sum_{j=2}^{i-1} (d_{+x}^j - d_{-x}^j), & \text{otherwise},
    \end{cases}
\end{equation}
which summarises the left and right turns of previous amino acids to determine which $x$-coordinate the amino acid has.
The locality of the position equations is lower than for the turn equations; the minus sign inside the summation in $x_i$ cancels out the highest-local term resulting in, at most next-to-highest-local terms in the 3D-case, we get:
\begin{align}
    d_{+x}^j - d_{-x}^j =& \ (1-q_{3j-4}) q_{3j-5}q_{3j-3} \nonumber \\
    &- (1-q_{3j-3}) q_{3j-5}q_{3j-4} \nonumber \\
    =& \ q_{3j-3}q_{3j-5} - q_{3j-3}q_{3j-4}q_{3j-5} \nonumber \\
    &- q_{3j-4}q_{3j-5} + q_{3j-3}q_{3j-4}q_{3j-5} \nonumber \\
    =& \ q_{3j-3}q_{3j-5} - q_{3j-4}q_{3j-5},
\end{align}
which also holds for $y_i$ and $z_i$. This is repeated in the 2D case where two-local equations for the position are reduced to one-local. The position equations have the locality of $D-1$ in \binary{} and \bubin{} encodings and a locality of one in the \oh{} encoding.

\paragraph*{Distance between amino acids: }
Moreover, we need equations that give the distance between two amino acids $j$ and $k$:
\begin{equation}
    D_{ij} = (x_i - x_j)^2 + (y_i - y_j)^2 + (z_i - z_j)^2.
    \label{app:eq:distaa}
\end{equation}
All terms in the distance equation are $2(D-1)$-local for the \binary{} and \bubin{} encodings, and two for the \oh{} encoding, as all other ordered terms cancel out. The number of interactions in the distance equation will be the sum of all possible pairs of the terms in the position equations. The terms for the maximal distance $\numaminoacids$ will contain the other distance terms, and the maximal number of terms is thus $\binom{\numaminoacids}{2}$. 

\paragraph*{Avoiding overlap.}
Auxiliary qubits are introduced to enforce the bounds on the distance function
\begin{equation}
    0 \leq D_{ij} \leq (i-j)^2,
\end{equation}
and to ensure no amino acids overlap
\begin{equation}
    D_{ij} \not = 0, \text{ if } i>j+3,
\end{equation}
thereby enforcing the inequality constraints
\begin{equation}
    D_{ij} \geq 1.
\end{equation}
The total number of auxiliary qubits needed to encode the information of the distance between amino acids for all amino acid pairs in the \binary{} encoding is
\begin{equation}
    \numqubits_{\text{aux}}^{\text{dist}} = \sum ^{\numaminoacids-5}_{i=0}\sum ^{\numaminoacids-1}_{j=i+4}\ceil{2 u \left(|1+i-j|\bmod 2\right)},
    \label{eq:ancolap}
\end{equation}
where the number of qubits $u$ to encode the information will differ between the encodings according to Table~\ref{tab:infostore}.

\begin{table}[hbp]
    \caption{\label{tab:infostore} Qubits needed to encode the
    information on the distance between two amino acids for each encoding. That is, the variable $u$ in Eq.~(\ref{eq:ancolap}).} 
\begin{ruledtabular}
\begin{tabular}{LL}
        \text{Encoding} & \text{Qubits ($u$)} \\ \hline 
        \text{\oh{}} & (j-i) \\ 
        \text{\binary{}} & \log_2(j-i) \\
        \text{\bubin{}} & \ceil{(j-i)/g} \ceil{\log_2{(g+1)} }\\
        \end{tabular}
\end{ruledtabular}
\end{table}

We introduce the slack variables
\begin{equation}
    \alpha_{ij} = \sum_{j=0}^{\mu_{ij}-1} q_{p_{ij}+j}2^{\mu_{ij}-1-j}
    \label{app:eq:slack}
\end{equation}
where $\mu_{ij}$ is the number of qubits needed to store the slack variable, and $p_{ij}$ is a pointer
\begin{equation}
    p_{ij} = (D\numaminoacids-8) + \sum_{u=0}^i \sum_{n=u+4}^{\numaminoacids-1}\mu_{un} - \sum_{m=j}^{\numaminoacids-1} \mu_{im},
\end{equation}
that will always point to qubits $q_{(D\numaminoacids-7),\dots, \numqubits_{\text{auxiliary}}}$.
The final Hamiltonian to avoid the overlap
\begin{align}
    H_{\text{overlap}} = \lambda_{\text{overlap}} &\sum_{i=0}^{\numaminoacids-5} \sum_{j=i+4}^{\numaminoacids-1} \left( |1+j-i| \bmod 2 \right) \nonumber \\
    &\times \left( 2^{\mu_{ij}} - D_{ij} - \alpha_{ij} \right)^2,
    \label{eq:app:overlap}
\end{align}
where $\mu_{ij}$ is an integer. Hence, we calculate the needed qubits by counting the terms linked to the following operators except the first one
\begin{align}
    \left( 2^{\mu_{ij}} - D_{ij} - \alpha_{ij} \right)^2 =& \ 2^{\mu_{ij}2} - 2^{\mu_{ij}+1} + D_{ij}^2 + \alpha_{ij}^2  \nonumber \\ &+ 2D_{ij}\alpha_{ij} - 2^{\mu_{ij}+1}\alpha_{ij}.
    \label{eq:app:overlapterms}
\end{align}
See Table~\ref{app:tab:interactionturn} for an overview of the $k$-locality and interacting qubits associated with the previous operators.
The number of interactions in the term $D_{ij}^2$, is all combinations of connecting two amino acids, and then all combinations of connecting these combinations two and two:
\begin{equation}
    \binom{\binom{\numaminoacids}{2}}{2} = \frac{1}{8} (\numaminoacids^4 -2\numaminoacids^3-\numaminoacids^2+2\numaminoacids)
    \propto \numaminoacids^4 ,
\end{equation}
and the locality is $4(D-1)$ for the \binary{} and \bubin{} encodings, and four for the \oh{} encoding. The number of interactions in the $\alpha_{ij}^2$ is all combinations of connecting the auxiliary qubits $\numqubits_{\text{aux}}^{\text{dist}}$ pairwise. The number of interactions in the term $2D_{ij}\alpha_{ij}$ is the number of interactions in $D_{ij}$ times the number of terms in $\alpha_{ij}$, which is equal to $\numqubits_{\text{aux}}^{\text{dist}}$. The last term $2^{\mu_{ij}+1}\alpha_{ij}$ has locality one.

\begin{table*}[htbp]
    \caption{\label{app:tab:interactionturn} The number of interactions and type of interaction for each operator in the double sum in Eq.~(\ref{eq:app:overlapterms}). Here, we denote the dimensionality of the lattice $D$, the number of amino acids $\numaminoacids$, the number of auxiliary qubits for encoding the distance $\numqubits_{\text{aux}}^{\text{dist}}$, the equation for distance between two amino acids $D_{ij}$ from Eq.(\ref{app:eq:distaa}), the slack variables' value $\alpha_{ij}$ from Eq.~(\ref{app:eq:slack}), and the number of qubits needed to store the slack variables $\mu_{ij}$.}
    \begin{ruledtabular}
    \begin{tabular}{LLLLL}
        \text{Operator} & \text{Interactions} & \text{Interacting qubits} & \text{$k$-locality \binary{} and \bubin{}} & \text{$k$-locality \oh{}}\\
        \hline 2^{\mu_{ij}2} - 2^{\mu_{ij}+1} & 0 & - & - & -\\
        \hline D_{ij}^2 & \dbinom{\tbinom{\numaminoacids}{2}}{2} & q_{0,\dots,(D\numaminoacids-8)} & 4(D-1) & 4\\
        \hline \alpha_{ij}^2 & \binom{\numqubits_{\text{aux}}^{\text{dist}}}{2} & q_{(D\numaminoacids-7), \dots, \numqubits_{\text{aux}}^{\text{dist}}} & 2 & 2\\
        \hline 2D_{ij}\alpha_{ij} & \binom{\numaminoacids}{2}\numqubits_{\text{aux}}^{\text{dist}} & q_{0, \dots, \numqubits_{\text{aux}}^{\text{dist}}} & 2(D-1)+1 & 3\\
        \hline -2^{\mu_{ij}+1}\alpha_{ij} & \numqubits_{\text{aux}}^{\text{dist}} & q_{(D\numaminoacids-7), \dots, \numqubits_{\text{aux}}^{\text{dist}}} & 1 & 1\\
    \end{tabular}
    \end{ruledtabular}
\end{table*}

\paragraph*{Pairwise amino acid interaction.}
The complexity of the Hamiltonian $H_{pair}$ in the turn-based model arises from the need for encoding distance information between amino acids to calculate their lattice interactions. This requires circuit-based calculations, demanding additional qubits to track interactions between specific amino acids ($j$ and $k$) on the lattice.
\begin{equation}
    \omega_{jk} = 
    \begin{cases}
      1, \text{ if } D_{jk}=1\\
      0, \text{ otherwise},
    \end{cases}
\end{equation}
and the number of auxiliary qubits needed are
\begin{equation}
    \numqubits_{\text{aux}}^{\text{pair}} = \sum_{j=0}^{\numaminoacids-4} \sum_{k=j+3}^{\numaminoacids-1} \left[ (|j-k|) \bmod 2 \right].
\end{equation}
The auxiliary qubits can then be used with an energy matrix (either MJ- or HP-potential) $P_{jk}$, and the expression on the pairwise Hamiltonian is formulated as 
\begin{align}
    H_{pair} =& \sum_{j=0}^{\numaminoacids-4} \sum_{k=j+3}^{\numaminoacids-1} \left( |j-k| \bmod 2 \right) \omega_{jk}P_{jk}(2-D_{jk}) \nonumber\\
    =& \sum_{j=0}^{\numaminoacids-4} \sum_{k=j+3}^{\numaminoacids-1} \left( |j-k| \bmod 2 \right) P_{jk}(2\omega_{jk}-D_{jk}\omega_{jk}).
\end{align}
The number of interactions to build the pair Hamiltonian is thereby given by the number of terms in $D_{ij}$ times the number of auxiliary qubits
\begin{equation}
    \numterms^{\text{pair}} = \binom{\numaminoacids}{2}\numqubits_{\text{aux}}^{\text{pair}}.
\end{equation}

\paragraph*{Total number of interactions.}
We get the total interaction for the turn-based lattice model
\begin{align}
    \numterms^{\text{total}} = & \ (D\numaminoacids-8)c_1 \nonumber \\
    +& \binom{\binom{\numaminoacids}{2}}{2} + \binom{\numqubits_{\text{aux}}^{\text{dist}}}{2} + \binom{\numaminoacids}{2}\numqubits_{\text{aux}}^{\text{dist}} \nonumber \\
    +& \binom{\numaminoacids}{2}\numqubits_{\text{aux}}^{\text{pair}},
\end{align}
where the constant $c_1$ is dependent on the lattice dimension and is small in comparison to the other terms.

\paragraph*{The total number of qubits.}
The total number of qubits is given by
\begin{equation}
    \numqubits_{\text{total}} = \numqubits_{\text{conf}} + \numqubits_{\text{aux}}^{\text{dist}} + \numqubits_{\text{aux}}^{\text{pair}} 
\end{equation}
For the \binary{} encoding this is  
\begin{align}
    \numqubits_{\text{total}}^{\text{\binary{}}} =& \ (D\numaminoacids-c_1) \nonumber \\
    +& \sum_{i=0}^{\numaminoacids-5} \sum_{j=i+4}^{\numaminoacids-1} \ceil{2 \log_2{(j-i)}} \left( |1+j-i| \bmod 2 \right) \nonumber \\
    +& \sum_{j=0}^{\numaminoacids-4} \sum_{k=j+3}^{\numaminoacids-1} \left( |j-k| \bmod 2 \right),
\end{align}
but here, we changed the order of $i$ and $j$ in the logarithm, as $j$ is always larger than $i$, and we cannot take the logarithm of a negative value.
The total number of qubits needed for the \oh{} encoding is then
\begin{align}
    \numqubits_{\text{total}}^{\text{\oh{}}} =& \ 2D\numaminoacids \nonumber \\
    +& \sum_{i=0}^{\numaminoacids-5} \sum_{j=i+4}^{\numaminoacids-1} \ceil{2 (j-i)} \left( |1+j-i| \bmod 2 \right) \nonumber \\
    +& \sum_{j=0}^{\numaminoacids-4} \sum_{k=j+3}^{\numaminoacids-1} \left( |j-k| \bmod 2 \right).
\end{align}
Lastly, the total number of qubits needed for the \bubin{} encoding is
\begin{align}
    \numqubits_{\text{total}}^{\text{\bubin{}}} =& \ \ceil{\log_2{(g+1)}}\ceil{\numaminoacids D/g}\nonumber \\
    +& \sum_{i=0}^{\numaminoacids-5} \sum_{j=i+4}^{\numaminoacids-1} 2 \ceil{(j-i)/g} \nonumber \\
    & \ceil{\log_2{(g+1)}} \left( |1+j-i| \bmod 2 \right) \nonumber \\
    +& \sum_{j=0}^{\numaminoacids-4} \sum_{k=j+3}^{\numaminoacids-1} \left( |j-k| \bmod 2 \right).
\end{align}